\documentclass[12pt]{iopart}

\usepackage{iopams}
\usepackage{color,amssymb,graphicx,pstricks,epsfig}
\newcommand{\nc}{\newcommand}
\nc{\nit}{\noindent} \nc{\nn}{\nonumber} \nc{\D}{\partial}
\nc{\DX}{\partial_X} \nc{\Av}[1]{\langle {#1} \rangle}
\nc{\AAv}[1]{\left\langle {#1} \right\rangle}
\nc{\diff}[2]{\frac{d #1}{d #2}} \nc{\diffn}[3]{\frac{d^{#3} #1}{d
{#2}^{#3}}} \nc{\pdiff}[2]{\frac{\partial #1}{\partial #2}}
\nc{\pdiffn}[3]{\frac{\partial^{#3} #1}{\partial{#2}^{#3}}}
\nc{\abs}[1] {\lvert #1 \rvert} \nc{\cE}{{\cal E}} \nc{\cF}{{\cal
F}} \nc{\cV}{{\cal V}} \nc{\cQ}{{\cal Q}} \nc{\cU}{{\cal U}}
\nc{\UU}{{\cal U}} \nc{\cGin}{{\cal G}_{\rm in}} \nc{\cGout}{{\cal
G}_{\rm out}} \nc{\cO}{{\cal O}} \nc{\Lav}{{\cal L}_{\rm av}}
\nc{\cL}{{\cal L}} \nc{\cB}{{\cal B}} \nc{\cZ}{{\cal Z}}
\nc{\cP}{{\cal P}} \nc{\cM}{{\cal M}} \nc{\vD}{{\vec D}}
\nc{\vE}{{\vec E}} \nc{\vB}{{\vec B}} \nc{\vH}{{\vec H}}
\nc{\ty}{{\tilde y}} \nc{\xoe}{{x\over\varepsilon}} \nc{\psit}{{
\tilde{\psi} } } \nc{\tpsi}{{\Phi}} \nc{\tu}{{\tilde u}}
\nc{\tV}{{\tilde V}} \nc{\tVhat}{\hat{\tilde V}} \nc{\bx}{{\bf x}}
\nc{\bk}{{\bf k}} \nc{\bX}{{\bf X}} \nc{\bXYZ}{{\bf XYZ}}
\nc{\bY}{{\bf Y}} \nc{\bZ}{{\bf Z}} \nc{\bF}{{\bf F}}
\nc{\bS}{{\bf S}} \nc{\dV}{{\delta V}} \nc{\dVN}{{\delta V_N}}
\nc{\dVNv}{{\delta \check{V}_N}} \nc{\dVt}{{\delta\tilde{V}}}
\nc{\dv}{{\delta v}} \nc{\dE}{{\delta E}} \nc{\dPsi}{{\delta\Psi}}
\nc{\pp}{\perp} \nc{\order}{{\cal O}} \nc{\Eplus}{E_+}
\nc{\Eminus}{E_-}
\nc{\Epm}{E_\pm}
\nc{\Vav}{V_{\rm av}}
\nc{\Rin}{R_{\rm in}}
\nc{\Rout}{R_{\rm out}}
\nc{\eplus}{e_+}
\nc{\eminus}{e_-}
\nc{\epm}{e_\pm}
\nc{\eps}{\epsilon}
\nc{\half}{{1\over2}}
\nc{\veps}{\varepsilon}
\nc{\vnabla}{{\vec\nabla}}
\nc{\G}{\Gamma}
\nc{\w}{\omega}
\nc{\mh}{h}
\nc{\mg}{g}
\nc{\sgn}{{\rm sgn}}
\nc{\vphi}{\varphi}
\nc{\tlambda}{\tilde\lambda}

\nc{\g}{\gamma}

\newtheorem{theo}{Theorem}[section]

\newtheorem{prop}{Proposition}[section]
\newtheorem{lem}{Lemma}[section]
\newtheorem{cor}{Corollary}[section]

\newtheorem{rmk}{Remark}[section]

\nc{\la}{\langle} \nc{\ra}{\rangle}
\nc{\infint}{\int_{-\infty}^{\infty}} \nc{\halfwidth}{6.5cm}
\nc{\figwidth}{10cm} \nc{\essspec}{\sigma_{\rm ess}}
\nc{\sqrtE}{\mu} \nc{\resolv}{R} \nc{\proofend}{$\square$}
\nc{\localize}{\chi} \nc{\localizehat}{\hat{\chi}}
\nc{\localizeinv}{\chi^{-1}} \nc{\localizet}{\tilde{\chi}}
\nc{\localizethat}{\hat{\localizet}}
\nc{\localizetinv}{\tilde{\chi}^{-1}} \nc{\mloc}{M}
\nc{\mtloc}{\tilde{M}} \nc{\smoothinv}{\langle D \rangle}
\nc{\smoothinvhat}[1][\xi]{\langle #1 \rangle}
\nc{\smooth}{\langle D \rangle^{-1}}
\nc{\smoothhat}[1][\xi]{\langle #1 \rangle^{-1}} \nc{\TR}{T_R}
\nc{\TRl}{T_{R,l}} \nc{\TRE}{T_R} \nc{\TRlE}{T_{R,l}}
\nc{\dTRE}{\delta\TRE} \nc{\TV}[1][V]{T_{#1}} \nc{\SR}{S_R}
\nc{\SV}[1][V]{S_{#1}} \nc{\twopin}{(2\pi)^n}
\nc{\twopiminusn}{(2\pi)^{-n}} \nc{\msmooth}{m}
\nc{\resD}{\mathcal{L}} \nc{\bessker}{G} \nc{\remker}{\mathcal{R}}
\nc{\TRI}{A_{\rm I}} \nc{\TRII}{A_{\rm II}} \nc{\TRIIa}{A_{\rm
II}^{(a)}} \nc{\TRIIb}{A_{\rm II}^{(b)}} \nc{\TRIII}{A_{\rm III}}
\nc{\genBessel}{\mathcal{J}} \nc{\bddcomm}{\mathcal{C}}
\nc{\Fhat}{\hat{F}} \nc{\fhat}{\hat{f}} \nc{\ghat}{\hat{g}}
\nc{\FM}{F_2^{(1)}}
\nc{\xiunder}{\underline{\xi}}
\nc{\xiover}{\overline{\xi}}
\nc{\Psizerobar}{\overline{\Psi}_0}
\nc{\psizerobar}{\overline{\psi}_0}
\nc{\Vt}{\tilde{V}}
\nc{\Psit}{\tilde{\Psi}}
\nc{\Et}{\tilde{E}}
\nc{\dEcoef}{C_{dE}}
\nc{\Etwohomog}{E_2^{\rm(homog)}}
\nc{\dEone}{\dE^{(1)}}
\nc{\dEtwo}{\dE^{(2)}}
\nc{\dEtwomain}{N_2^{\rm{(PT,main)}}}
\nc{\dEtworem}{N_2^{\rm{(PT,rem)}}}
\nc{\fbar}{\overline{f}}
\nc{\leftfn}{f^{(L)}}
\nc{\rightfn}{f^{(R)}}
\nc{\partialwaveG}{G}
\nc{\supp}{\operatorname{supp}}
\nc{\rt}{\tilde{r}}

\nc{\PU}{\chi} \nc{\gt}{\tilde{g}} \nc{\indicator}{\boldsymbol{1}}
\nc{\Ai}{\mathop{\mathrm{Ai}}\nolimits}
\nc{\thetat}{\tilde{\theta}}\nc{\gms}{V}\nc{\NN}{ {\tilde N} }
\nc{\II}{\operatorname{I_1}} \nc{\III}{C_V}

\begin{document}

\title[]{Analytic theory of narrow lattice solitons}

\author{Y.~Sivan}
\address{School of Physics and Astronomy, Tel Aviv University,
Tel Aviv 69978, Israel} \ead{yonatans@post.tau.ac.il}
\author{G.~Fibich}
\address{School of Mathematical Sciences, Tel Aviv University,
Tel Aviv 69978, Israel}
\author{N.K.~Efremidis}
\address{Department of Applied Mathematics, University of Crete,
71409 Heraklion, Crete, Greece} \address{School of Electrical and
Computer Engineering, National Technical University of Athens,
Athens 15773, Greece}
\author{S.~Bar-Ad}
\address{School of Physics and Astronomy, Tel Aviv University, Tel
Aviv 69978, Israel}

\begin{abstract}
The profiles of narrow lattice solitons are calculated analytically
using perturbation analysis. A stability analysis shows that
solitons centered at a lattice (potential) maximum or saddle point
are unstable, as they drift toward the nearest lattice minimum. This
instability can, however, be so weak that the soliton is
``mathematically unstable'' but ``physically stable''. Stability of
solitons centered at a lattice minimum depends on the dimension of
the problem and on the nonlinearity. In the subcritical and
supercritical cases, the lattice does not affect the stability,
leaving the solitons stable and unstable, respectively. In contrast,
in the critical case (e.g., a cubic nonlinearity in two transverse
dimensions), the lattice stabilizes the (previously unstable)
solitons. The stability in this case can be so weak, however, that
the soliton is ``mathematically stable'' but ``physically
unstable''.
\end{abstract}


\maketitle

\section{Introduction}\label{eq:intro}
Solitons are localized waves that propagate in nonlinear media where
dispersion and/or diffraction are present. They appear in various
fields of physics such as nonlinear optics, Bose-Einstein
Condensates (BEC), plasma physics, solid state physics and water
waves. The dynamics of solitons is modeled by the Nonlinear
Schr\"odinger equation (NLS) in the context of nonlinear optics
which is also known as the Gross-Pitaevskii (GP) equation in the
context of BEC.

In the study of stability of solitons in a homogeneous medium, it is
useful to consider the $d$-dimensional focusing NLS
\begin{equation}\label{eq:nls-p}
iA_z(z,{\bf x})\ =\ - \nabla^2 A - |A|^{p-1}A,
\end{equation}
where $z$ is the longitudinal coordinate, ${\bf x} = (x_1, \dots,
x_d)$ are the coordinates in the transverse plane, $\nabla^2 =
\partial_{{x_1}{x_1}} + \cdots + \partial_{{x_d}{x_d}}$ is the
Laplacian operator and the nonlinearity is focusing with exponent
$p>1$. In optics, the $z$ variable in Eq.~(\ref{eq:nls-p}) is
normalized by $2L_{diff}$, where $L_{diff}$ is the diffraction
(Rayleigh) length and the $x_j$ variables are normalized by the
input beam radius.

We delineate several cases for the NLS~(\ref{eq:nls-p}):
\begin{eqnarray}
0\ <\ &p - 1\ <\ \frac{4}{d},\ \ \ {\rm the\ subcritical\ case},\nonumber\\
&p - 1\ =\ \frac{4}{d},\ \ \ {\rm the\ critical\ case,}\nonumber\\
&p - 1\ >\ \frac{4}{d},\ \ \ {\rm the\ supercritical\ case}.
\label{subcritsuper}
\end{eqnarray}
In the subcritical case, the solitary waves $A=e^{i \nu z}u_\nu({\bf
x})$ of the NLS~(\ref{eq:nls-p}) are stable, while in the critical
and supercritical cases the solitary waves of the
NLS~(\ref{eq:nls-p}) are unstable. The profile of a stable solitary
wave experiences only minor changes under small perturbations as it
propagates. On the other hand, unstable solitary waves can change
dramatically due to the effect of an infinitesimal perturbation. For
the NLS~(\ref{eq:nls-p}), unstable solitary waves either collapse
after propagating a finite distance, or diffract as $z$ goes to
infinity~\cite{We:1989,We:1983}.

Solitons have been thoroughly studied in view of their potential
application in optical communications and switching devices (in
nonlinear optics) or in quantum information science (in BEC). Recent
advances in fabrication and experimental methods now make possible
the realization of transparent materials with spatially varying,
high contrast dielectric properties. Such materials have various
all-optical signal processing applications in optical
communications, see {\it e.g.}~\cite{chris-nature2003,Joann}. In
this case, the solitons are usually called {\em lattice solitons}.
Specifically, by a proper design of the dielectric properties of the
medium, it may be possible to avoid the blowup/diffraction
instability in the critical and supercritical cases and to obtain
stable propagation of laser beams in those
structures~\cite{Ostrovskaya-Kivshar-rev,Segev_lattice_solitons_review,aceve-pre1996,Aceves1}.
Thus, there is considerable interest in understanding the
propagation of light in modulated media.

Most studies of such media have considered {\it linear lattices}
(potentials). In this case, the equation of propagation is
\begin{equation}
iA_z(z,{\bf x}) = -\nabla^2 A - |A|^{p-1}A + \gms(N {\bf x}_{lat})
A, \label{eq:NLS_linear_lattice_dp}
\end{equation}
where ${\bf x}_{lat} = (x_1, \dots,x_{d_{lat}})$ are the lattice
coordinates, $1 \le d_{lat} \le d$ is the lattice dimension and
$1/N$ is the characteristic length-scale of change of the lattice.
For example, if the lattice is periodic, then $N$ is the lattice
period. In the context of nonlinear optics, linear potentials are
created by modulating the linear refractive index $n_0$ in space. If
the modulation/potential is periodic, such structures are called
{\em waveguide arrays} or {\em photonic lattices}. In the context of
BEC, the corresponding Gross-Pitaevskii equation accounts for the
interaction of the atoms with a magnetic trap or, in the case of a
periodic optical lattice, with interfering laser beams,
see~\cite{Abdullaev_BEC_review,Konotop_Brazhnyi_rev} and references
therein.

Solitary waves of the NLS~(\ref{eq:NLS_linear_lattice_dp}) with a
general linear potential were studied in~\cite{Oh:89,RW:88}, to
name a few of the earlier studies. Recently, many studies
considered periodic potentials. Theoretical and numerical studies
of solitons of the NLS/GP equation were done for a periodic
potential in
one~\cite{chris-ol1988,Kivshar-pra2003-lattice-soliton-profile,efrem-pra2003,Pelinovsky_theory_1D_l_ms},
two~\cite{efrem-prl2003,yang-ol2003,ZiadY} and
three~\cite{mihal-pra2005,mihal-3d-optics} dimensions.
Experimental realization of these solitons were obtained in
one-dimensional waveguide arrays~\cite{eisen-prl1998} and in
two-dimensional optically induced photonic lattices in
photorefractive
media~\cite{efrem-pre2002pr,fleis-prl2003,Segev-etal,Neshev_OL,pertc-ol2004femto}.
Some studies also involved lattices whose dimensionality is
smaller than the spatial dimension, i.e., $d_{lat} < d$ (see
e.g.~\cite{Aceves1,Malomed_low_d}) and in media with a quintic
nonlinearity (see~\cite{alfimov_p5} and references therein).

Generally speaking, it was found that for some lattice types and
propagation constants $\nu$, the lattice can prevent the collapse
and stabilize the solitons in the critical and supercritical
cases. However, the possibility that these stable solitons can
collapse under a sufficiently large perturbation was not mentioned
in previous studies.

A detailed study of stability (and collapse) of solitons in a {\em
nonlinear} lattice, i.e.,
\begin{equation}
iA_z(z,{\bf x}) = - \nabla^2 A - \gms(N {\bf x}_{lat})|A|^{p-1}A,
\label{eq:NLS_nllinear_lattice_dp}
\end{equation}
was done in~\cite{NLS_NL_MS_1D,NLS_NL_MS_2D}. In these studies it
was shown that the soliton profile and (in)stability properties
strongly depend on whether it is wider than, of the same order of,
or narrower than the lattice period. Specifically, it has been shown
that the same nonlinear lattice may stabilize beams of a certain
width while destabilizing beams of a different width. Hence, any
study of the stability of lattice solitons should take into account
the (relative) soliton width.

In this paper, we conduct a systematic study of the stability and
instability dynamics of solitons in linear lattices which are {\em
narrow} with respect to the lattice period. The fact that the
solitons are narrow imply that there is a small non-dimensional
parameter $\NN$, see Eq.~(\ref{eq:NN_intro}). This allows us to
employ perturbation methods and to compute the soliton profile and
related quantities (soliton power, perturbed zero-eigenvalues
$\lambda_{0,j}^{(N)}$, see below) asymptotically.

In nonlinear optics, typical lattice periods are of the order of
several microns and typical input beam sizes are not smaller than
this
period~\cite{eisen-prl1998,Segev-etal,Dima-PRL-2003,pertc-prl2004,Segev_lattice_solitons_review}.
Hence, typically, the input beam sizes are not small compared with
the lattice period. However, if the beam undergoes collapse, the
beam can become much narrower than the lattice period. In BEC, the
standard magnetic traps are significantly wider than the size of the
condensate. Hence, the narrow beams limit is of physical relevance.
From a theoretical point of view, the limit of narrow beams
corresponds to the semi-classical limit of the nonlinear
Schr\"odinger equation
\begin{equation}
ihA_z(z,{\bf x}) = -h^2\nabla^2 A - |A|^{p-1}A + \gms({\bf
x}_{lat})A, \quad \quad h \longrightarrow 0,
\label{eq:NLS_linear_lattice_dp_semiclassical}
\end{equation}
see e.g.,~\cite{Oh:89,Lin_Wei}. Moreover, as discussed in
Section~\ref{sec:summary}, in many cases, the results for narrow
beams hold also for beams of $\cO(1)$ width.

The paper is organized as follows: In Section~\ref{sec:Definition},
we present various physical models in nonlinear optics and in BEC
where Eq.~(\ref{eq:NLS_linear_lattice_dp}) arises. In
Section~\ref{sec:solution}, the equation for lattice soliton is
derived. It is shown that the soliton width is given by a {\em
single} parameter
\begin{equation} \label{eq:NN_intro}
\NN = \frac{N}{\sqrt{\eta}} \ll 1, \quad \quad \eta = \nu + \gms(0),
\end{equation}
where $V(0)$ is the potential at the soliton center. Therefore, the
limit $\nu \to \infty$ analyzed in~\cite{Reika-05-gen-pot}, and the
limit $N \to 0$ analyzed in~\cite{Lin_Wei}, are in fact the same
limit. It is well known that narrow solitons of a periodic lattice
are found deep inside the ``semi-infinite gap'' of the linear
problem, away from the first band of the allowed
solutions~\cite{efrem-pra2003}, i.e., for~$\nu \to \infty$. Indeed,
in this case $\NN \to 0$. However, from this argument it is not
clear how large should~$\nu$ be in order for the soliton to be
narrow. This information is given by the parameter~$\NN$, which is
thus, a more informative parameter than the propagation
constant~$\nu$. Moreover, the parameter~$\NN$ includes also the
effect of the lattice strength on the width and reflects the fact
that as~$V(0)$ increases, the beam confinement increases, hence the
beam becomes narrower~\footnote{Note, however, that
expression~(\ref{eq:NN}) for the beam relative width is only valid
for narrow beams.}.

In Section~\ref{sec:solution}, we also use perturbation analysis to
calculate the profile of narrow lattice solitons for any dimension
$d$, lattice dimensionality $d_{lat}$ and nonlinearity exponent $p$.
As can be expected, this calculation shows that the soliton profile
depends only on the local properties of the lattice, rather than on
the full lattice structure. Hence, {\em our study is relevant to any
slowly varying lattice}, regardless of its long-scale properties. To
simplify the notation, we mostly consider lattices that are aligned
in the directions of the Cartesian axes. In this case, the lattice
can be expanded as
\begin{eqnarray}\label{eq:VN_NN_intro}
V(N {\bf x}_{lat}) = V(0) + \eta \left(N^2 \sum_{j=1}^{d_{lat}}
v_{jj} x^2_j + \cO(\NN^4)\right).
\end{eqnarray}
Our results are valid, however, to any linear lattice, see
Remark~\ref{rmk:axes}.

In Section~\ref{sec:stability}, we analyze the stability of narrow
lattice solitons. We first present the two conditions for stability
of lattice solitons in Theorem~\ref{theo:stability-nls}. The first
condition, known as the Vakhitov-Kolokolov
condition~\cite{vakhi-rqe1973} or the {\em slope
condition}~\cite{We:86}, is that the power (or $L_2$ norm) of the
soliton should increase with $\nu$. Using the results of the
perturbation analysis, we show in Section~\ref{sec:slope} that to
leading order, the power of a narrow lattice soliton is equivalent
to the power of a soliton in a homogeneous medium, and that the
change in the power due to the lattice scales as
$\NN^2$.~\footnote{For comparison, the change in the power due to a
{\em nonlinear} lattice is $\cO(\NN^2)$ in the subcritical and
supercritical cases but $\cO(\NN^4)$ in the critical
case~\cite{NLS_NL_MS_1D,FW:03}. } In particular, the lattice causes
the power to decrease (increase) for lattice solitons centered at a
lattice minimum (maximum). In addition, the power curve slope is
more positive (negative) for lattice solitons centered at a lattice
minimum (maximum). Since in a homogeneous medium the slope has an
$\cO(1)$ magnitude in the subcritical and supercritical cases, the
small change of the slope by the lattice does not affect the sign of
the slope. Accordingly, the slope condition remains satisfied in the
subcritical case and violated in the supercritical case. In the
critical case, the slope in a homogeneous medium is zero. As a
result, the $\cO(\NN^2)$ change in the power by the lattice leads to
a positive (negative) slope for lattice solitons centered at a
lattice minimum (maximum). Hence, the slope condition is satisfied
for narrow lattice solitons centered at a lattice minimum, but is
``even more'' violated for lattice solitons centered at a lattice
maximum.

The second condition for stability of narrow lattice solitons is the
{\em spectral condition}~\cite{ManosG}, and it involves the number
of negative eigenvalues of the linearized operator
$L^{(N)}_{+,\nu}$, see Eq.~(\ref{eq:L_+N}). In
Section~\ref{sec:spectral}, we first show that the spectral
condition is violated if and only if the lattice causes some of the
zero eigenvalues of the homogeneous medium linearized operator
$L_{+,\nu}$ (see Eq.~(\ref{eq:L+nu})) to become negative. Then, we
use a perturbation analysis to show that the values of the perturbed
zero eigenvalues $\lambda_{0,j}^{(N)}$ are given by
\begin{eqnarray}\label{} \nn
\lambda_{0,j}^{(N)} &=& \left\{
\begin{array}{llll}
\delta v_{jj} N^2 + \cO(\NN^4), & & & j = 1, \dots, d_{lat}, \\
0, & & & j = d_{lat}+1, \dots, d,
\end{array}
\right.
\end{eqnarray}
where
\begin{eqnarray}
\delta &=& \frac{p(2-d) + 2+ d}{p-1}, \nn \label{eq:}
\end{eqnarray}
see Lemma~\ref{lem:ev}. This calculation shows that the eigenvalues
become positive (negative) for solitons centered at a lattice
minimum (maximum). Hence, the {\em spectral} condition is satisfied
(violated) for solitons centered at a lattice minimum (maximum).
This calculation generalizes the result of Oh in the one-dimensional
cubic case~\cite{Oh:89} to any dimension $d$, any lattice dimension
$d_{lat}$ and any nonlinearity exponent $p$.

In order to test the validity of the analytical formula for
$\lambda_{0,j}^{(N)}$, we also compute these eigenvalues
numerically. For $d \ge 2$, the matrix that represents the
linearized operator $L_{+,\nu}^{(N)}$ is very large. As a result,
standard numerical schemes (e.g., Matlab's {\tt eig} or {\tt eigs})
usually fail to compute its eigenvalues. In order to overcome this
numerical difficulty, we use a numerical scheme which is based on
the Arnoldi algorithm, see~\ref{app:ev}. While in this study we
``only'' use this scheme to verify the validity of the analytical
approximation of the eigenvalue, we note that in the case of
non-narrow lattice solitons, the eigenvalue cannot be computed
analytically, and the only way to check the spectral condition is
numerically. Moreover, this numerical scheme can be used in similar
eigenvalue problems in which large matrices are involved.

Combining the results of Sections~\ref{sec:slope}
and~\ref{sec:spectral}, we show in
Section~\ref{sub:stability_results}
(Proposition~\ref{prop:narrow_bs_stability}) that {\em in the
subcritical and critical cases, narrow lattice solitons are stable
when centered at a lattice minimum, and unstable when centered at a
lattice maximum or at a saddle point. In the supercritical case,
narrow lattice solitons are unstable at both lattice maxima and
minima}. 

Proposition~\ref{prop:narrow_bs_stability} specifies when the two
conditions for stability are violated. It does not, however,
describe the resulting instability dynamics. 
The relations between the condition which is violated and the
instability dynamics were observed
in~\cite{NLS_NL_MS_1D,NLS_NL_MS_2D} for a {\em nonlinear} lattice
and in~\cite{delta_pot_complete} for a linear delta-potential to be
as follows:
\begin{enumerate}
  \item if the slope is negative, the soliton width can undergo significant
changes. In the critical and supercritical cases, this {\em width
instability} can result in collapse. In the subcritical case, this
width instability can ``only'' result
in a ``finite-width'' instability, i.e., the soliton width can decrease 
substantially, but not to zero.
  \item When the spectral condition is violated, the
solitons undergo a {\em drift} instability, i.e., the soliton
drifts away from the lattice maximum towards the nearest lattice
minimum.
  \item When both conditions for stability are violated, a combination of a
width instability and a drift instability can be observed.
\end{enumerate}
In the case of narrow lattice solitons, the slope is always positive
in the subcritical case. Hence, the instability due to a negative
slope is a blowup instability and not a ``finite-width''
instability. Furthermore, in Section~\ref{sub:charac} we {\em prove}
that when the spectral condition is violated (i.e., if the soliton
is centered at a lattice maximum or saddle point), narrow lattice
solitons undergo a drift instability, i.e., they move away from
their initial location at an exponential drift-rate. In contrast,
solitons centered near a lattice minimum (for which the spectral
condition is satisfied) undergo small oscillations around the
lattice minimum. The above observations on the condition leading to
instability and the type of instability dynamics are summarized in
Table~\ref{tab:stability}.

\begin{table}
\begin{center}
\begin{tabular}{|l||c|c|}
\hline & lattice minimum & lattice maximum \\ \hline \hline
Subcritical & {\em Stability} & Instability$^\dagger$ (drift)\\
\hline Critical &  {\em Stability} & Instability$^{*,\dagger}$ (blowup+drift)\\
\hline Supercritical & Instability$^*$ (blowup) & Instability$^{*,\dagger}$ (blowup+drift)\\
\hline
\end{tabular}

\end{center} \caption{Stability of narrow lattice solitons. Condition leading to instability is marked by $^*$ for a
failure to satisfy the slope condition and by $^\dagger$ for a
failure to satisfy the spectral condition. In the case of
instability, its dynamics is indicated in parentheses.
}\label{tab:stability}
\end{table}

In Section~\ref{sec:qualtitative}, we study the dynamics of solitons
in the two cases where the small effect of the lattice changes the
stability. As observed in~\cite{NLS_NL_MS_1D,NLS_NL_MS_2D}, in such
cases, it is important to study both stability and instability {\em
quantitatively}. In Section~\ref{sub:slope}, we discuss the {\em
strength of the stabilization} induced by the lattice for solitons
centered at a lattice minimum in the critical case. To do so, we use
the concept of the {\em stability region}, i.e., the region in
function space of initial conditions around the soliton profile that
lead to a stable propagation. As in the case of a {\em nonlinear}
lattice~\cite{NLS_NL_MS_1D,NLS_NL_MS_2D}, our results indicate that
the $\cO(\NN^2)$ small slope of the power curve implies that the
stability region is $\cO(\NN^2)$ small~\footnote{In the case of a
nonlinear lattice, the slope, hence the size of the stability
region, is $\cO(\NN^4)$ small, implying an even weaker
stability~\cite{NLS_NL_MS_1D,FW:03}. }. Therefore, although the two
conditions for stability are satisfied, these solitons can become
unstable under extremely small perturbations. Practically, this
means that in the critical case, ``mathematically'' stable solutions
can be ``physically'' unstable i.e., become unstable under typical
physical perturbations. We illustrate these results using two
standard types of lattices: A sinusoidal potential, which is typical
in photorefractive materials~\cite{efrem-pre2002pr,Segev-etal} and
in BEC~\cite{tromb-prl2001} and a Kronig-Penney step lattice
(periodic array of finite potential wells)~\cite{carr-pra2001},
which is typical for manufactured slab waveguide arrays, see
e.g.,~\cite{chris-ol1988,eisen-prl1998,pertc-prl2004}. We study
numerically the stability of solitons under random perturbations
that either increase or decrease the total power of the soliton and
observe that narrow lattice solitons are ``mathematically'' stable
but ``physically'' unstable. The stability is particularly weak for
Kronig-Penney lattice solitons, for which the slope is exponentially
small. In addition, we observe that when the perturbation is
sufficiently ``non-small'', both the sinusoidal and KP (stable)
lattice solitons can undergo a {\em blowup instability}. This shows
that in the absence of translation invariance, stability and blowup
can coexist in NLS equations~\cite{NLS_NL_MS_1D,NLS_NL_MS_2D,FM:01}.

In Section~\ref{sub:drift}, we show that the opposite scenario is
also possible, i.e., ``mathematically unstable'' solitons can be
``physically stable''. This occurs for subcritical narrow lattice
solitons centered at a lattice maximum, which are unstable due to a
violation of the spectral condition
(Proposition~\ref{prop:narrow_bs_stability}). We show that the drift
rate is exponential in $\bigg(-\lambda_0^{(N)}\bigg)^{1/2}$.
Therefore, narrow solitons, for which $\lambda_0^{(N)}$ is
$\cO(N^2)$ small, experience very slow drift and can thus be
``stable'' for the distances/times in experimental setups. In
particular, we observe that the Kronig-Penney lattice soliton drifts
much more slowly than the sinusoidal lattice soliton of the same
width. Section~\ref{sec:summary} concludes with some concluding
remarks.

\section{Physical models}\label{sec:Definition}
We consider the d dimensional NLS
equation~(\ref{eq:NLS_linear_lattice_dp}) with a linear lattice in
$d_{lat}$~dimensions ($1 \le d_{lat} \le d$). This model describes
numerous physical configurations. For example, beam propagation in
a Kerr slab waveguide with a lattice is modeled by
\begin{equation}\label{eq:NLS_linear_lattice_1d}
i A_z(z,x) = - A_{xx} - |A|^2A + \gms(Nx) A.
\end{equation}
In this case, $p=3$, $d = d_{lat} = 1$, ${\bf x} = {\bf x_{lat}} =
x$ and $\gms = \gms(Nx)$, see
e.g.,~\cite{Malomed_1D_lattice_solitons,efrem-pra2003,eisen-prl1998}.
Beam propagation in bulk Kerr medium with a two-dimensional
lattice is modeled by
\begin{equation}\label{eq:NLS_linear_lattice_2d}
i A_z(z,x,y) = - \nabla^2 A - |A|^2A + \gms A.
\end{equation}
In this case, $p=3$, $d = 2$ and ${\bf x} = (x,y)$. If $\gms =
\gms(Nx,Ny)$, then $d_{lat} = 2$, and ${\bf x}_{lat} = (x,y)$, see
e.g.,~\cite{yang-ol2003,ZiadY,efrem-prl2003}; if $\gms = \gms(Nx)$
then $d_{lat} = 1$, and ${\bf x}_{lat} = x$. In the latter case,
the dimension of the lattice $d_{lat}$ is smaller by one from the
dimension of the transverse space $d$, see
e.g.,~\cite{Aceves1,NLS_NL_MS_2D,Malomed_low_d}.

Propagation of ultrashort pulses in a slab waveguide is modeled by
\begin{equation}\label{eq:NLS_linear_lattice_t1}
i A_z(z,x,t) = - A_{xx} + \beta_2 A_{tt} - |A|^2A + \gms(Nx) A,
\end{equation}
where $\beta_2$ is the group velocity dispersion (GVD) parameter.
In the case of anomalous dispersion ($\beta_2 < 0$), the time
coordinate $t$ is effectively an additional transverse dimension.
Then, Eq.~(\ref{eq:NLS_linear_lattice_t1}) corresponds to
Eq.~(\ref{eq:NLS_linear_lattice_dp}) with $p=3$, $d = 2$, ${\bf x}
= (x,t)$, $d_{lat} = 1$ and ${\bf x}_{lat} = x$, so the dimension
of the lattice $d_{lat}$ is smaller by one from the dimension of
the transverse space $d$, see e.g.,~\cite{Aceves1,NLS_NL_MS_2D}.
Similarly, propagation of ultrashort pulses in a 2D optical
lattice is modeled by
\begin{equation}\label{eq:NLS_linear_lattice_t2}
i A_z(z,x,y,t) = - \nabla^2 A + \beta A_{tt} - |A|^2A + \gms A,
\end{equation}
which for $\beta < 0$ corresponds to
Eq.~(\ref{eq:NLS_linear_lattice_dp}) with $p=3$, $d = 3$ and ${\bf
x} = (x,y,t)$. If $\gms = \gms(Nx,Ny)$, then $d_{lat} = 2$, and
${\bf x}_{lat} = (x,y)$~\cite{mihal-3d-optics}; if $\gms =
\gms(Nx)$ then $d_{lat} = 1$, and ${\bf x}_{lat} = x$.

The linear lattice $\gms$ in Eq.~(\ref{eq:NLS_linear_lattice_dp})
varies in the transverse coordinates but not in $z$. In some
applications, the lattice varies in the direction of
propagation~$z$. 
Such problems, however, will not be studied in this paper.

Eq.~(\ref{eq:NLS_linear_lattice_dp}) also models the dynamics of
Bose-Einstein condensates (BEC) with a negative scattering length.
In this case, $z$ is replaced with $t$. In BEC, typically ${\bf x}
= (x,y,z)$, i.e., $d=3$, but under certain conditions the cases $d
= 1$ and $d = 2$ are also of physical interest, see
e.g.,~\cite{1D_BEC_3D_trap,Dimensional_reduction_markowich}. The
exponent $p$ is usually equal to $3$ but can also be equal to $5$,
see~\cite{alfimov_p5} and references therein. In the BEC context,
both a parabolic potential and a periodic potential appear in the
experimental setups~\cite{Abdullaev_BEC_review}.

\section{Narrow lattice solitons}\label{sec:solution}
We look for lattice solitons, which are of solutions of
Eq.~(\ref{eq:NLS_linear_lattice_dp}) of the form
\begin{equation}
A(z,{\bf x}) = e^{i \nu z}u_\nu^{(N)}({\bf x}), \quad \quad \nu
> 0, \label{eq:uN-def}
\end{equation}
where $u_\nu^{(N)}$ is the solution of
\begin{equation}
\nabla^2 u_\nu^{(N)}({\bf x}) + \big(u_\nu^{(N)}\big)^p - [\nu +
\gms(N {\bf x}_{lat})] u_\nu^{(N)} = 0. \label{eq:uN-eqn}
\end{equation}
We consider lattices which are symmetric with respect to a critical
point ${\bf x}_{lat}^{(0)}$ of the lattice $\gms$. Hence, the
soliton maximal amplitude is attained at ${\bf
x}_{lat}^{(0)}$~\cite{Grossi}. The boundary conditions for
Eq.~(\ref{eq:uN-eqn}) are $\nabla u_\nu^{(N)}({\bf x}_{lat}^{(0)}) =
0$ and $u_\nu^{(N)}(\infty) = 0$. Without loss of generality, we set
${\bf x}_{lat}^{(0)} = 0$.

We study solutions of Eq.~(\ref{eq:uN-eqn}) which are narrow with
respect to the lattice characteristic length-scale. A priori, the
relative width of a lattice depends on the lattice strength, the
lattice period (or characteristic length) $1/N$ and the propagation
constant~$\nu$. We now show that in the case of narrow solitons, one
can rescale Eq.~(\ref{eq:uN-eqn}) to a form where the relative width
of the beam is given by a {\em single} parameter $\NN$. In order to
achieve that, we define
\begin{equation}\label{eq:eta}
\eta = \nu + \gms(0), \quad \quad \NN = N / \sqrt{\eta}, \quad
\quad u_\nu^{(N)}({\bf x}) = \eta^{\frac{1}{p-1}}
u_{\NN}(\sqrt{\eta}{\bf x}).
\end{equation}
Then, Eq.~(\ref{eq:uN-eqn}) becomes
\begin{equation}
\nabla^2 u_\NN(\tilde {\bf x}) + u_\NN^p - [1 + \tilde \gms(\NN
\tilde {\bf x}_{lat} )] u_\NN = 0, \quad \nabla u_\NN(0), \quad
u_\NN(\infty)=0,  \label{eq:uNN-eqn}
\end{equation}
where
\begin{equation}\label{eq:scaling_add}
\tilde {\bf x} = \sqrt{\eta} {\bf x}, \quad \quad \tilde {\bf
x}_{lat} = \sqrt{\eta} {\bf x}_{lat}, \quad \quad \tilde \gms (\NN
\tilde {\bf x}_{lat}) = \frac{\gms(\NN \tilde {\bf x}_{lat}) -
\gms(0)}{\eta}.
\end{equation}
When $\NN \ll 1$, we can expand the solution of
Eq.~(\ref{eq:uNN-eqn}) as a power series of $\NN^2$, i.e.,
\begin{equation}\label{eq:expansion}
u_\NN(\tilde {\bf x}) = \UU(|\tilde {\bf x}|) + \NN^2 g(\tilde
{\bf x}) + \cO(\NN^4),
\end{equation}
where $\UU$ is the positive, radially-symmetric ground-state
solution of
\begin{equation}
\nabla^2 \UU(|\tilde {\bf x}|) + \UU^p - \UU = 0.
\label{eq:UU-eqn}
\end{equation}
Similarly, since $\tilde \gms(0) = 0$ and $\nabla \tilde \gms(0) =
0$, the potential $\tilde \gms(\NN \tilde {\bf x}_{lat})$ can be
expanded for $\NN \ll 1$ as
\begin{equation}
\tilde \gms(\NN \tilde {\bf x}_{lat}) = \NN^2 \tilde \gms_2(\tilde
{\bf x}_{lat}) + \cO(\NN^4), \label{eq:lattice-series}
\end{equation}
where
\begin{equation} \label{eq:v2}
\tilde \gms_2(\tilde {\bf x}_{lat}) = \sum_{j,k=1}^{d_{lat}}
v_{jk} \tilde x_j \tilde x_k, \quad \quad v_{jk} = \frac{1}{2}
 \frac{\partial^2 \tilde \gms({\bf y}_{lat})}{\D {y_j}
 \partial {y_k} }\bigg|_{{\bf y}_{lat} = 0},
\end{equation}
is the first non-vanishing term in the Taylor expansion of $\tilde
\gms$ which represents the local curvature of the lattice at the
soliton center. In particular, $\tilde \gms_2(\tilde {\bf
x}_{lat}) \ge 0$ ($ \le 0$) for lattice solitons centered at a
lattice minimum (maximum).

\begin{rmk}\label{rmk:axes}
In order to simplify the presentation, we assume that the
principle axes of the lattice identify with the Cartesian axes
$\{\hat{e}_1,\dots,\hat{e}_{d_{lat}}\}$. In this case, $v_{jk} =
0$ for $j \ne k$,
\begin{equation} \label{eq:v2_jj}
\tilde \gms_2(\tilde {\bf x}_{lat}) = \sum_{j=1}^{d_{lat}} v_{jj}
\tilde x_j^2,
\end{equation}
and
\begin{eqnarray}\label{eq:VN_NN}
V(N {\bf x}_{lat}) = V(0) + \eta \left(N^2 \sum_{j=1}^{d_{lat}}
v_{jj} x^2_j + \cO(\NN^4)\right),
\end{eqnarray}
see Eq.~(\ref{eq:VN_NN_ap}). However, all our results can be
immediately generalized to the case when the lattice is not aligned along the cartesian axes as follows. Since $v_{jk} = v_{kj}$, 
there exists a basis of vectors
$\{\hat{\eps}_1,\dots,\hat{\eps}_{d_{lat}}\}$ such that if $\tilde
{\bf x}_{lat} = \sum_{j=1}^{d_{lat}} \alpha_j \hat{\eps}_j$ then
\begin{equation}\label{eq:ujj}
\tilde V_2(\tilde {\bf x}_{lat}) = \sum_{j=1}^{d_{lat}} u_{jj}
\alpha_j^2.
\end{equation}
Therefore, in order to apply our results to the
lattice~(\ref{eq:v2}), one needs to replace $x_j$ by $\alpha_j$
and $v_{jj}$ by $u_{jj}$. See e.g.,
Remark~\ref{rmk:lattice_effect} and Remark~\ref{rmk:ev}.
\end{rmk}

Using a perturbation analysis similar to the one used
in~\cite{NLS_NL_MS_1D,FW:03}, we show that
\begin{lem}\label{lem:sol}
The solution of Eq.~(\ref{eq:uNN-eqn}) for $\NN \ll 1$ is given by
\begin{equation}\label{eq:uNN-sol}
u_\NN(\tilde {\bf x}) = \UU(|\tilde {\bf x}|) - \NN^2
L_+^{-1}\left( \tilde \gms_2(\tilde {\bf x}_{lat}) \UU \right) +
\cO(\NN^4),
\end{equation}
where $\UU$ is given by Eq.~(\ref{eq:UU-eqn}), $\tilde \gms_2$ is
given by Eq.~(\ref{eq:v2_jj}) and
\begin{equation}
L_+ = -\nabla^2_{\bf \tilde x} - p\UU^{p-1} + 1. \label{eq:L+}
\end{equation}
\end{lem}

\nit{\bf Proof:} See~\ref{ap:nls-proof}.

In the original variables, the expansion~(\ref{eq:uNN-sol})
becomes
\begin{eqnarray}\label{eq:u-nu-N-sol}\fl
u_\nu^{(N)}({\bf x}) &=& (\nu + V(0))^{\frac{1}{p-1}} \left[
\UU(\sqrt{\nu + V(0)}|{\bf x}|) + \cO(\NN^2) \right] =
\UU_\eta(|{\bf x}|) + \cO(\NN^2),
\end{eqnarray}
where $\UU_\eta = \eta^{\frac{1}{p-1}}\UU(\sqrt{\eta}|\tilde {\bf
x }|)$ is the solution of
\begin{equation}
\nabla^2 \UU_\eta(|{\bf x}|) + \UU_\eta^p - \eta \UU_\eta = 0.
\label{eq:UU-eta-eqn}
\end{equation}

This expansion shows that:
\begin{enumerate}
  \item To leading order, a (rescaled) narrow lattice soliton $u_\NN$ is given by the
rescaled homogeneous medium soliton $\UU$.
  \item The deviation of a narrow lattice soliton $u_\NN$ from $\UU$ is $\cO(\NN^2)$ small, even if the
lattice has $\cO(1)$ variations.
\end{enumerate}
The above results also show that the soliton relative width is
given by a {\em single} parameter~$\NN$:
\begin{prop}
Lattice solitons are narrow with respect to the lattice period if
\begin{equation}\label{eq:NN}
\NN = \frac{N}{\sqrt{\eta}} = \frac{N}{\sqrt{\nu + \gms(0)}} \ll
1.
\end{equation}
In this case,
\begin{equation}\label{eq:NN_words}
\NN = \frac{\mbox{soliton width}}{\mbox{lattice period}}.
\end{equation}
\end{prop}

\nit {\bf Proof:} By Eq.~(\ref{eq:uNN-sol}), when $\NN \ll 1$,
then $u_\NN$ has $\cO(1)$ width in ${\bf \tilde x}$. Hence, by
Eq.~(\ref{eq:u-nu-N-sol}), the width of $u_\nu^{(N)}({\bf x})$ in
${\bf x}$ is $\cO(1/\sqrt{\eta})$. Since that the lattice
length-scale/period is $1/N$, then the relative width of the
soliton is given by $\NN$. $\Box$

We emphasize that the expansion~(\ref{eq:uNN-sol}) applies to {\em
all} types of lattices so long as $v_{jj} \NN^2 \ll 1$.
Specifically, for a strong periodic lattice ($V \gg 1$), for which
the linear coupling between adjacent lattice sites is weak, the
result~(\ref{eq:u-nu-N-sol}) is still valid provided that $\NN$ is
small enough, i.e., for $\NN \ll v_{jj}^{-\frac{1}{2}}$. In that
case, the solution~(\ref{eq:u-nu-N-sol}) is the continuous analog of
the discrete solitons of the DNLS
model~\cite{chris-ol1988,eisen-prl1998,Bose_Eins_Exp1}.

\subsection{Effect of lattice type}\label{subsec:lat_type}
Lemma~\ref{lem:sol} shows that the effect of the lattice depends
on whether $\tilde \gms_2 \not\equiv 0$ or $\tilde \gms_2 \equiv
0$. When $\tilde \gms_2 \not\equiv 0$, then the lattice effect is
$\cO(\NN^2)$. This case corresponds to a parabolic lattice, a
sinusoidal lattice etc. However, when $\tilde \gms_2 \equiv 0$,
then $g({\bf x}) \equiv 0$ and the next-order term in the
expansion~(\ref{eq:expansion}) must be considered. In particular,
in the special case of a Kronig-Penney step lattice (see, e.g.,
Eq.~(\ref{eq:V_step_2D})), all derivatives of $\tilde \gms$ at the
soliton center ${\bf x}_{lat}^{(0)} = 0$ vanish. Therefore, the
difference between $u_\NN$ and $\UU$ will be exponentially small.

\subsection{Effect of lattice inhomogeneity on soliton profile}\label{subsec:lat_inhom}
In order to calculate the effect of the lattice on the soliton
profile, we note that
\begin{equation}\label{eq:N2correction}
L_+^{-1} \left( \tilde x_j^2 \UU \right) = \tilde x_j^2 S(|\tilde
{\bf x}|) + Q(|\tilde {\bf x}|),
\end{equation}
where $S$ and $Q$ are radial functions which are the solutions of
\begin{eqnarray}
L_+ S - \frac{4}{\tilde r} S' = \UU, \quad \quad L_+ Q = 2 S.
\label{eq:SQ}
\end{eqnarray}
Indeed, applying the operator $L_+$ to the right-hand-side of
Eq.~(\ref{eq:N2correction}) gives
\begin{eqnarray}\label{eq:Lp_g} \fl
L_+ \left( \tilde x_j^2 S(|\tilde {\bf x}|) + Q(|\tilde {\bf x}|)
\right)&=& \left(- \nabla^2_{\tilde {\bf x}} + 1 - p\UU^{p-1}
\right) \left(\tilde x_j^2 S(r) + Q(\tilde r) \right)
\nonumber \\
&=& \tilde x_j^2 \underbrace{\left( L_+ S(\tilde r) -
\frac{4}{\tilde r} S'(\tilde r) \right)}_{= \UU} -
\underbrace{\left(2 S(\tilde r) + L_+ Q(\tilde r)\right)}_{ = 0}
\nn = \tilde x_j^2 \UU.
\end{eqnarray}
Therefore, the $\cO(\NN^2)$ correction to the soliton profile due
to the lattice~(\ref{eq:v2_jj}) is given by, see
Eq.~(\ref{eq:uNN-sol}),
\begin{equation}\label{eq:N2correction2}\fl
u_\NN - \UU \sim - \NN^2 L_+^{-1} \left( V_2(\tilde {\bf x}_{lat})
\UU \right) = - \NN^2 \Big( \underbrace{ S(|\tilde {\bf x}|)
\sum_{j=1}^{d_{lat}} v_{jj} \tilde x_j^2}_{\mbox{anisotropic}} \
\underbrace{ + Q(|\tilde {\bf x}|) \sum_{j=1}^{d_{lat}}
v_{jj}}_{\mbox{isotropic}}\Big).
\end{equation}
Thus, the variation of the lattice in the direction $x_j$ has an
isotropic effect through $Q(|\tilde {\bf x}|)$ and an anisotropic
effect in the direction $x_j$ through $S(|\tilde {\bf x}|)$.

\begin{rmk}\label{rmk:lattice_effect}
If $\tilde V_2$ is given by the lattice~(\ref{eq:v2}), then, the
$\cO(\NN^2)$ correction to the soliton profile is given by
\begin{equation}\label{eq:N2correction3}\fl
u_\NN - \UU \sim - \NN^2 L_+^{-1} \left( V_2(\tilde {\bf x}_{lat})
\UU \right) = - \NN^2 \left(S(|\tilde {\bf x}|)
\sum_{j=1}^{d_{lat}} u_{jj} \tilde \alpha_j^2 + Q(|\tilde {\bf
x}|) \sum_{j=1}^{d_{lat}} u_{jj}\right).
\end{equation}
\end{rmk}

\section{Stability and instability of lattice solitons}\label{sec:stability}

Eq.~(\ref{eq:u-nu-N-sol}) implies that narrow lattice solitons
$u_\nu^{(N)}$ are positive. The conditions for stability and
instability of positive lattice solitons are as
follows~(\cite{NLS_NL_MS_1D}, and see
also~\cite{We:85,We:86,ManosG}):
\begin{theo}\label{theo:stability-nls}
Let $u^{(N)}_\nu$ be a positive solution of Eq.~(\ref{eq:uN-eqn}), 
let $\mathcal{P}_\nu^{(N)} \equiv
\int \left(u_\nu^{(N)}\right)^2 d{\bf x}$ be the power of
$u_\nu^{(N)}$, and let $n_-(L^{(N)}_{+,\nu})$ be the number of
negative eigenvalues of the linearized operator
\begin{equation}
L_{+,\nu}^{(N)} = -\nabla^2 + \nu - p\big(u^{(N)}_\nu({\bf
x})\big)^{p-1} + \gms(N {\bf x}_{lat}). \label{eq:L_+N}
\end{equation}
Then, the lattice soliton $A(z,{\bf x}) = e^{i \nu
z}u_\nu^{(N)}({\bf x})$ is
\begin{enumerate}
  \item an orbitally {\em stable} solution of
Eq.~(\ref{eq:NLS_linear_lattice_dp}) if
\begin{enumerate}
\item $\D_\nu \mathcal{P}^{(N)}_\nu > 0$ (slope
condition), and
\item $n_-(L^{(N)}_{+,\nu}) = 1$ (spectral condition).
\end{enumerate}
\item an orbitally {\em unstable} solution of
Eq.~(\ref{eq:NLS_linear_lattice_dp}) if
\begin{enumerate}
\item $\D_\nu \mathcal{P}^{(N)}_\nu < 0$, or
\item $n_-(L^{(N)}_{+,\nu}) > 1$.
\end{enumerate}
\end{enumerate}
\end{theo}
In what follows, we use the expansion~(\ref{eq:uNN-sol}) to
determine whether the two conditions in
Theorem~\ref{theo:stability-nls} are satisfied ,and consequently
determine the stability of narrow lattice solitons.

\subsection{Slope condition}\label{sec:slope}
We can use the expansion~(\ref{eq:uNN-sol}) to calculate the power
of narrow lattice solitons:
\begin{lem}\label{lem:power}
The power of narrow lattice solitons ($\NN \ll 1$) is given by
\begin{eqnarray}
\mathcal{P}_\nu^{(N)} &=& \left(\nu + \gms(0)\right)^{\frac{4 -
d(p-1)}{2(p-1)}}\left( \mathcal{P}_{\nu=1} - \III \NN^2
\sum_{j=1}^{d_{lat}} v_{jj} + \cO(\NN^4)\right),
\label{eq:un_power2_res}
\end{eqnarray}
where $\mathcal{P}_{\nu=1} = \int \left|\UU\right|^2 d\tilde {\bf
x}$, $\UU$ is the positive solution of Eq.~(\ref{eq:UU-eqn}), and
\begin{eqnarray}\label{eq:C_V}
\III = \frac{2p - 6 + d p - d}{2d(p-1)}
\int | \tilde {\bf x}|^2 \UU^2 d \tilde {\bf x}
\end{eqnarray}
is a constant independent of $N$ and $\nu$.
\end{lem}

\nit {\bf Proof:} See~\ref{app:power-proof}.

Eq.~(\ref{eq:un_power2_res}) shows that, in a similar manner to
its effect on the soliton profile, when $\tilde \gms_2 \not\equiv
0$ (e.g., in the case of a sinusoidal or parabolic lattice), the
lattice has an $\cO(\NN^2)$ small effect on the soliton power,
even if the lattice itself is not weak. In light of
Section~\ref{subsec:lat_type}, in the case of a Kronig-Penney step
lattice, the effect of the lattice on the power is exponentially
small in $\NN$.

>From Eq.~(\ref{eq:C_V}) it also follows that $\III
> 0$ for $p > 1 + \frac{4}{2+d}$. In particular, for $p = 3$, $\III =
\frac{1}{2}\int \tilde r^2 \UU^2 d \tilde {\bf x} > 0$. Thus,
\begin{cor}
If $\tilde V_2 \not\equiv 0$, the lattice causes the power to
decrease (increase) for lattice solitons centered at a lattice
minimum (maximum) for any $p > 1 + \frac{4}{2+d}$, and in
particular, for a Kerr nonlinearity $p=3$.
\end{cor}

\begin{center}\begin{figure}[ht!]
\includegraphics[width=16cm]{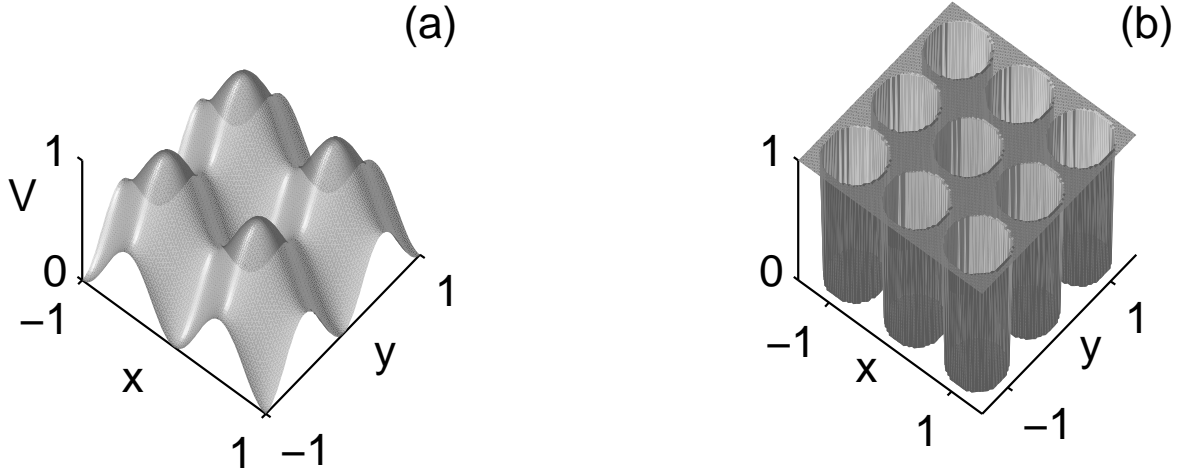} \caption{(a) sinusoidal lattice~(\ref{eq:V_sin_2D}) (b) Kronig-Penney lattice~(\ref{eq:V_step_2D}). }
\label{fig:lattices}
\end{figure}
\end{center}

In order to demonstrate the result of Lemma~\ref{lem:power}, we
solve Eq.~(\ref{eq:uNN-eqn}) numerically with $d = d_{lat} = 2$,
${\bf x} = {\bf x}_{lat} = (x,y)$ and $p=3$. For convenience, the
numerical results shown here are presented for $\eta = 1$ (so that
$N = \NN$, $u_\nu^{(N)} = u_\NN$) and for $\gms(0) = 0$ (so that
$\gms = \tilde \gms$). We study two different two-dimensional
lattices with a periodic square topology:
\begin{enumerate}
  \item A 2D {\em sinusoidal lattice}
given by
\begin{equation} \label{eq:V_sin_2D}
\gms(Nx,Ny) = \pm \frac{1}{2}\left(\sin^2(\pi {N}x) + \sin^2(\pi
{N} y)\right).
\end{equation}
  \item A 2D {\em Kronig-Penney lattice} that consists
of an array of primitive cells of size $[-1/N\ 1/N]\times[-1/N\
1/N]$, each consisting of circular waveguide with abrupt index
change between 0 and $1$, i.e.,
\begin{equation}
\gms (N x,N y) = \left\{
\begin{array}{ll}
0, & \sqrt{x^2+y^2} < \frac{N}{N_0}, \quad \quad N_0 \cong 2.45\\
\pm 1, & \mathrm{otherwise. }
\end{array}
\right. \label{eq:V_step_2D}
\end{equation}
\end{enumerate}
In both cases, the plus/minus sign corresponds to a lattice with a
minimum/maximum at ${\bf x}_c = 0$, respectively. The parameters
of these lattices were chosen so that both lattices have a period
1, mean value $1/2$ and vary from $0$ to $\pm1$. The lattices are
shown in Fig.~\ref{fig:lattices} for a lattice with a minimum at
${\bf x}_c = 0$. Note that both lattices are anisotropic in $r =
\sqrt{x^2 + y^2}$, and thus, require a full $2$-dimensional
treatment. Moreover, since the 2D cubic NLS is critical,
$\mathcal{P}_{\nu=1} = \mathcal{P}_{cr} \cong 11.7$, where
$\mathcal{P}_{cr}$ is the critical power for collapse in a
homogeneous Kerr medium.

\begin{center}\begin{figure}[ht!]
\includegraphics[width=6cm]{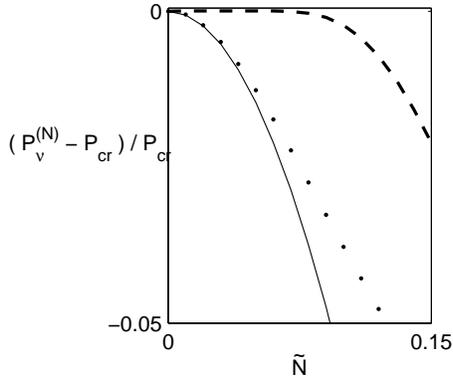} \caption{ Relative power deviation from $\mathcal{P}_{\nu=1} = \mathcal{P}_{cr}$ for narrow
sinusoidal (dots) and Kronig-Penney (dashes) lattice solitons
centered at a lattice minimum. The analytical
prediction~(\ref{eq:un_power_example}) for the sinusoidal lattice is
shown by a solid line. } \label{fig:power_2D}
\end{figure}
\end{center}

In Fig.~\ref{fig:power_2D}, we show the power of narrow lattice
solitons centered at a lattice minimum for both lattices. For $0 \le
N \le 0.1$ there is good agreement between the numerically
calculated value of the power of the sinusoidal lattice solitons and
the analytical approximation~\footnote{The agreement between the
analytic result~(\ref{eq:un_power2_res}) and the numerics is good
``only'' for relatively small values of $\NN$ because of the large
curvature ($\sum_{j=1}^2 v_{jj} = 4\pi^2$) of the lattice which
translates into a large coefficient of the $\NN^2$ term in
Eq.~(\ref{eq:un_power_example}). Indeed, we verified that for
smaller values of $\sum_{j=1}^2 v_{jj}$, the agreement between the
analytic result~(\ref{eq:un_power2_res}) and the numerics extends to
larger values of $\NN$.}
\begin{eqnarray}\fl
\mathcal{P}_\nu^{(N)} &=& \mathcal{P}_{\nu=1} - \III \NN^2
\sum_{j=1}^{d_{lat}} v_{jj} + \cO(\NN^4) \cong 11.7 - 6.94 \cdot
2(2\pi^2) \NN \cong 11.7 - 273.8 \NN^2,
\label{eq:un_power_example}
\end{eqnarray}
which is derived from Lemma~\ref{lem:power}. In particular, the
effect of the lattice on the power of the narrow lattice solitons
is much more pronounced in the case of a sinusoidal lattice than
in the case of a Kronig-Penney lattice.

The sign of the slope follows directly from
Eq.~(\ref{eq:un_power2_res}):

\begin{cor}\label{cor:slope}
Let $\NN \ll 1$. Then, the slope $\D_\nu P_\nu^{(N)}$ is positive
in the subcritical case ($p < 1+4/d$) and negative in the
supercritical case ($p > 1+4/d$). In the critical case ($p =
1+4/d$), the slope is positive for narrow lattice solitons
centered at a lattice minimum and negative for narrow lattice
solitons centered at a lattice maximum.
\end{cor}

\nit {\bf Proof}: In the subcritical and supercritical cases, the
slope is given by
\begin{eqnarray} \label{eq:non_critical_slope} \fl
\D_\nu \mathcal{P}_\nu^{(N)} &= \D_\nu \left( \left(\nu +
\gms(0)\right)^{\frac{4 - d(p-1)}{2(p-1)}} \left[
\mathcal{P}_{\nu=1} + \cO\left(\frac{N^2}{\nu +
V(0)}\right)\right]\right) \\ &= \left(\nu + \gms(0)\right)^{\frac{4
- d(p-1)}{2(p-1)} - 1} \left[ \mathcal{P}_{\nu=1} +
\cO\left(\frac{N^2}{\nu + V(0)}\right)\right] \sim \left(\nu +
\gms(0)\right)^{\frac{4 - d(p-1)}{2(p-1)} - 1} \mathcal{P}_{\nu=1}.
\nonumber
\end{eqnarray}
Therefore, in the subcritical case, the slope is positive while in
the supercritical case, the slope is negative. Note that in these
cases, the lattice does not affect the sign of the slope.

In the critical case, the first term in
Eq.~(\ref{eq:non_critical_slope}) vanishes and the slope is
determined by the $\cO(\NN^2)$ correction in
Eq.~(\ref{eq:un_power2_res}), i.e.,
\begin{eqnarray}\fl
\D_\nu \mathcal{P}_\nu^{(N)} &=& 0 - \III \frac{\partial
\NN^2}{\partial \nu}\sum_{j=1}^{d_{lat}} v_{jj} + \cO(\NN^4) = 2
\III \frac{\NN^2}{\nu + \gms(0)}\sum_{j=1}^{d_{lat}} v_{jj} +
\cO(\NN^4), \label{eq:critical_slope}
\end{eqnarray}
where we also used Eq.~(\ref{eq:eta}). By Eq.~(\ref{eq:C_V}), in
the critical case $\III = \frac{1}{d} \int \tilde r^2 \UU^2 d
\tilde {\bf x} > 0$, which completes the proof. $\Box$.

We thus conclude that {\em although the lattice has a small effect
on the profile of narrow lattice solitons, in the critical case,
this small effect determines the sign of the power slope and hence,
the stability (but see Section~\ref{sub:slope}).}

\subsection{Spectral condition} \label{sec:spectral}
As noted is Section~\ref{sec:stability}, lattice solitons are
stable only if in addition to the slope condition, they also
satisfy the spectral condition. In the absence of a lattice (i.e.,
for $\gms \equiv 0$), the linearized operator $L^{(N)}_{+,\nu}$
reduces to $L_{+,\nu}$ which is given by
\begin{equation}
L_{+,\nu} = -\nabla^2 - p\UU_\nu^{p-1} + \nu, \label{eq:L+nu}
\end{equation}
where $\UU_\nu = \nu^{\frac{1}{p-1}}\UU(\sqrt{\nu}|\tilde {\bf
x}|)$ and $\UU$ is given by Eq.~(\ref{eq:UU-eqn}). The spectrum of
$L_{+,\nu}$ consists of~\cite{We:85}:
\begin{enumerate}
  \item A negative eigenvalue $\lambda_{min}$ and a corresponding even and positive eigenfunction $f_{\nu,min}$. In~\cite{Oh:89}, Oh shows that for $d=1$ and $p=3$, $\lambda_{min} = -3\nu$ and $f_{\nu,min} = \UU^2$. More generally, we observe that for any value of $p$ and $d$,
  $$\lambda_{min} = - \frac{1}{4}\left(p - 1\right)\left(p + 3\right) \nu, \quad \quad f_{\nu,min} = \UU_\nu^{\frac{p+1}{2}}.
  $$
  \item A zero eigenvalue $\lambda_0$ of multiplicity $d$ with the
corresponding eigenfunctions
\begin{equation}\label{eq:wj0}
f_{\nu,j}({\bf x}) = \frac{\partial \UU_\nu}{\partial x_j} =
\frac{x_j}{{\bf |x|}} \UU_\nu'({\bf |x|}), \quad \quad j = 1,
\dots,d.
\end{equation}
  \item A positive continuous spectrum $[\nu,\infty)$.
\end{enumerate}
Thus, in a homogeneous medium the spectral condition is satisfied.
In the presence of a linear lattice, the perturbed smallest
eigenvalue $\lambda_{min}^{(N)}$ remains negative. The continuous
spectrum develops a band structure, but remains positive. Moreover,
for $d_{lat} < j \le d$, the $j$th perturbed zero eigenvalue remains
at zero with the corresponding eigenfunction~$\frac{\partial
u_\nu^{(N)}}{\partial x_j}$. Therefore, $L_{+,\nu}^{(N)}$ can attain
more than one negative eigenvalue only if at least one
$\lambda_{0,j}^{(N)}$ becomes negative for $1 \le j \le
d_{lat}$~\cite{NLS_NL_MS_1D}. Thus, in order to check if the
spectral condition is satisfied, we only need to compute the sign of
$\lambda_{0,j}^{(N)}$ for $1 \le j \le d_{lat}$.

For $d=1$, $p=3$ and a slowly varying parabolic potential, the value
of the perturbed zero eigenvalue $\lambda_0^{(N)} =
\lambda_{0,1}^{(N)}$ was computed by Oh~\cite{Oh:89}:~\footnote{The
formula given in~\cite{Oh:89} contains a minor error, since in pp.
29 of~\cite{Oh:89}, the $L_2$ norm of $\UU$ was used instead of the
$L_2$ norm of $\UU'$. }
\begin{equation}\label{eq:ev_Oh}
\lambda_0^{(N)} = 3 v_{jj} N^2 + \cO(N^3).
\end{equation}
A more general result on the value and sign of
$\lambda_{0,j}^{(N)}$ in the presence of a linear lattice for $d
\ge 2$ is not known to us. We now give an asymptotic formula for
$\lambda_{0,j}^{(N)}$ for narrow lattice solitons which
generalizes of the result of Oh to any dimension $d$, lattice
dimension $d_{lat}$ and nonlinearity $p$:
\begin{lem}\label{lem:ev}
Let $V$ be given by Eq.~(\ref{eq:VN_NN}), or equivalently, let
$\tilde V_2$ be given by Eq.~(\ref{eq:v2_jj}), and let $\NN \ll
1$. Then, the perturbed zero eigenvalues $\lambda_{0,j}^{(N)}$ of
the operator $L_{+,\nu}^{(N)}$ are given by
\begin{eqnarray}\label{eq:ev_expansion_res}
\lambda_{0,j}^{(N)} &=& \left\{
\begin{array}{llll}
\delta v_{jj} N^2 + \cO(\NN^4), & & & j = 1, \dots, d_{lat}, \\
0, & & & j = d_{lat}+1, \dots, d,
\end{array}
\right.
\end{eqnarray}
where
\begin{eqnarray}
\delta &=& \frac{p(2-d) + 2+ d}{p-1}. \label{eq:delta}
\end{eqnarray}
\end{lem}
\nit {\bf Proof}: See~\ref{app:ev}.

\begin{rmk}\label{rmk:ev} If $V$ has the general form~(\ref{eq:v2}),
then, Eq.~(\ref{eq:ev_expansion_res}) becomes
\begin{eqnarray}
\lambda_{0,j}^{(N)} &=& u_{jj} N^2 \delta + \cO(\NN^4), \quad \quad
j = 1, \dots, d_{lat}, \label{eq:ev_expansion_res_u}
\end{eqnarray}
and Eq.~(\ref{eq:delta}) remains unchanged.
\end{rmk}

\begin{prop}\label{prop:spectral_cond}
Let
\begin{equation}
\left\{
\begin{array}{llll}
1 < p, & & & d = 1,2\\
1 < p < \frac{d+2}{d-2}, & & & d > 2
\end{array}
\right. .\label{eq:p_spectral}
\end{equation}
Then, the spectral condition is satisfied for narrow lattice
solitons centered at a lattice minimum, and violated for narrow
lattice solitons centered at a lattice maximum.
\end{prop}

\nit {\bf Proof}: It is easy to verify that $\delta > 0$ if and
only if $p$ satisfies condition~(\ref{eq:p_spectral}). Thus,
Lemma~\ref{lem:ev} shows that
$$
sgn(\lambda_{0,j}^{(N)}) = sgn(v_{jj}).
$$
Consequently, the operator $L_{+,\nu}^{(N)}$ has one negative
eigenvalue ($\lambda_{0,j}^{(N)} > 0$) for a narrow lattice
soliton centered at a lattice minimum ($v_{jj} > 0$) and more than
one negative eigenvalue ($\lambda_{0,j}^{(N)} < 0$) for a narrow
lattice soliton centered at a lattice maximum ($v_{jj}<0$).
$\Box$.

We note that values of $p$ for which
condition~(\ref{eq:p_spectral}) is satisfied include all the
physically relevant cases of $d=1,2,3$ and $p = 3,5$.

\begin{center}\begin{figure}[ht!]
\includegraphics[width=10 cm]{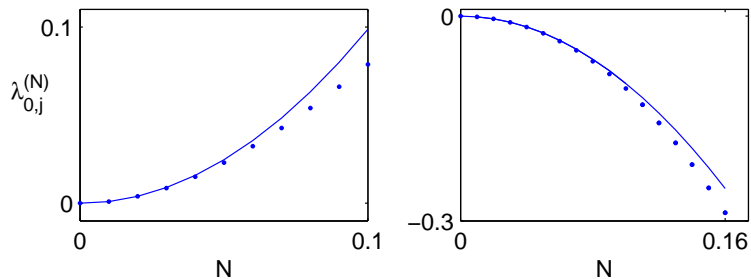}
\caption{ Eigenvalue $\lambda_{0,j}^{(N)}$ ($j=1,2$) of the operator
$L_{+,\nu}^{(N)}$ as a function of $\NN$ for the
lattice~(\ref{eq:V_sin_2D}) and a for soliton centered at a lattice
minimum (left) and at a lattice maximum (right). For $\NN \ll 1$,
there is a good agreement between the numerically calculated
eigenvalue of the discretized operator $L_{+,\nu}^{(N)}$ (dots) and
the analytical approximation~(\ref{eq:evs_narrow_2D}) (solid line).
} \label{fig:evs}
\end{figure}
\end{center}

To demonstrate the results of Lemma~\ref{lem:ev}, we consider the
case of $d = d_{lat} = 2$, $p=3$ and the
lattice~(\ref{eq:V_sin_2D}). By Eq.~(\ref{eq:ev_expansion_res}),
\begin{eqnarray}
\lambda^{(N)}_{0,1} = \lambda^{(N)}_{0,2} \cong 2 v_{jj} N^2 = \pm
(2\pi)^2 N^2. \label{eq:evs_narrow_2D}
\end{eqnarray}
In order to confirm the validity of the
expansion~(\ref{eq:evs_narrow_2D}), we compute the eigenvalues of
the discretized operator $L_{+,\nu}^{(N)}$ for the
lattice~(\ref{eq:V_sin_2D}). In general, for $d \ge 2$, computation
of the eigenvalues of the discretized operator $L_{+,\nu}^{(N)}$
(using, e.g., Matlab's {\tt eig} or {\tt eigs}) fails to give
reliable solutions due to computer memory limitation. In order to
overcome this limitation, we used an improved numerical scheme based
on the Arnoldi algorithm (see~\ref{app:code}). In Fig.~\ref{fig:evs}
we see that indeed for $N \ll 1$, the asymptotic
expression~(\ref{eq:evs_narrow_2D}) for the eigenvalue is in
agreement with its numerically calculated value.

\subsection{Stability results}\label{sub:stability_results}
Now that we have determined when the slope and spectral conditions
are satisfied, we can characterize the stability of narrow lattice
solitons:
\begin{prop}\label{prop:narrow_bs_stability} Let $\NN \ll
1$, let $u_\nu^{(N)}$ be the solution of Eq.~(\ref{eq:uN-eqn}),
let $p$ satisfy conditions~(\ref{eq:p_spectral}) and let $V$ be
given by Eq.~(\ref{eq:VN_NN}). Then,
\begin{enumerate}
  \item If $u_\nu^{(N)}$ is centered at a lattice
maximum, then $u_\nu^{(N)} e^{i \nu z}$ is unstable.
  \item If $u_\nu^{(N)}$ is centered at a lattice minimum, then
$u_\nu^{(N)} e^{i \nu z}$ is stable in the subcritical and
critical cases $p \le 1+4/d$, and unstable in the supercritical
case $p > 1+4/d$.
\end{enumerate}
\end{prop}

\nit {\bf Proof}: Instability of narrow lattice solitons centered
at a lattice maximum follows from a violation of the spectral
condition (Proposition~\ref{prop:spectral_cond}). For narrow
lattice solitons centered at a lattice maximum the spectral
condition is satisfied (Proposition~\ref{prop:spectral_cond}) and
stability is determined by the slope condition. Hence, the
stability in the subcritical and critical cases and instability in
the supercritical case follow from Corollary~\ref{cor:slope}.
$\Box$.

Proposition~\ref{prop:narrow_bs_stability} refers only to solitons
centered at a lattice minimum or maximum. In some cases (e.g., in
studies of lattices with defects or surface/corner
solitons~\cite{Ablowitz_Ilan_irreg_lattices}), lattice solitons can
be centered at critical points of the lattice that are saddle
points. In these cases, by Lemma~\ref{lem:ev}, the narrow lattice
solitons are unstable since the spectral condition is violated.

\subsection{Instability dynamics}\label{sub:charac}
Proposition~\ref{prop:narrow_bs_stability} specifies the conditions
for which narrow lattice solitons are unstable. It does not,
however, describe the instability dynamics that occur when those
conditions are not met. As noted in the Introduction, in previous
studies~\cite{NLS_NL_MS_1D,NLS_NL_MS_2D,delta_pot_complete} it was
observed that if the slope is negative, the solitons undergo a {\em
width instability} and when the spectral condition is violated, the
solitons undergo a {\em drift} instability.

In the case of narrow lattice solitons we can {\em prove} that
violation of the spectral condition results in a drift instability
by monitoring the dynamics of the soliton center of mass:
\begin{lem}\label{lem:com}
Let $\Av{x_j}$ be the center of mass in the $x_j$ coordinate,
i.e.,
\begin{equation}\label{eq:com}
\Av{x_j} \equiv \frac{\int x_j |A|^2 d{\bf x}}{\int|A|^2 d{\bf
x}}.
\end{equation}
Then,
\begin{equation}
\left\{
\begin{array}{llll}
\Av{x_j(z)} \sim \Av{x_j(0)} \cos(\Omega z) +
\frac{\Av{\dot{x}_j(0)}}{\Omega} \sin(\Omega z), & & & v_{jj} > 0, \\
&&& \\
\Av{x_j(z)} \sim \Av{x_j(0)} \cosh(\Omega z) +
\frac{\Av{\dot{x}_j(0)}}{\Omega} \sinh(\Omega z), & & & v_{jj} < 0,
\end{array}
\right. \label{eq:com_dyn_sol}
\end{equation}
where
\begin{equation}\label{eq:Omega}
\Omega = 2 N \sqrt{d \eta |v_{jj}|},
\end{equation}
and $v_{jj}$ defined in Eq.~(\ref{eq:v2_jj}).
\end{lem} \nit {\bf Proof}: See~\ref{app:accel}.

Thus, if $v_{jj} > 0$, the center of mass $\Av{x_j}$ oscillates
around the lattice minimum. On the other hand, if $v_{jj} < 0$,
the center of mass moves away from the lattice maximum at an
exponential rate. This shows, in particular, that a soliton
centered at a saddle point is stable in the directions in which it
is centered at a lattice minimum and undergoes a drift instability
in the directions in which it is centered at a lattice maximum.

\section{Quantitative study of stability}\label{sec:qualtitative}
As noted, the lattice has a small $\cO(\NN^2)$ effect on the slope
and on the value of the perturbed near zero-eigenvalues of
$L_{+,\nu}^{(N)}$. Nevertheless, this small effect changes the
stability of solitons centered at a lattice maximum (which became
unstable) and of solitons centered at a lattice minimum in the
critical case (which become stable). As pointed out
in~\cite{NLS_NL_MS_1D,NLS_NL_MS_2D}, when a small effect changes
the stability, stability and instability needs also to be studied
quantitatively.

\subsection{``Mathematical'' stability vs. ``physical''
stability}\label{sub:slope}
Let us first consider narrow lattice
solitons centered at a lattice minimum in the critical case. In
this case, according to Proposition~\ref{prop:narrow_bs_stability}
the solitons are stable. However, as was shown
in~\cite{NLS_NL_MS_1D,NLS_NL_MS_2D}, satisfying the
``mathematical'' conditions for stability does not necessarily
``prevent'' the development of instabilities due to small
perturbations. In order to understand how this can happen, we
recall that Theorem~\ref{theo:stability-nls} ensures that there is
a {\em stability region} in the function space of initial
conditions around the soliton profile for which the solution
remains stable. However, {\em it does not say how large this
stability region is}. If the stability region is very narrow, the
solution is only stable under extremely small perturbations. In
this case, it is ``mathematically'' stable but ``physically
unstable'', i.e., it can become unstable under perturbations
present in an experimental setup. If, on the other hand, it is
also stable under perturbations comparable in magnitude to
perturbations in actual physical setups, one can say that it is
also ``physically stable''.

The distinction between ``mathematical stability'' and ``physical
stability'' is only important in the critical case where, in the
absence of the lattice, the slope is zero. Then, the slope (VK)
condition shows that these solitons are unstable and indeed, an
arbitrarily small perturbation can cause them either to undergo
diffraction or to collapse. The effect of a linear lattice on
narrow lattice solitons centered at a lattice minimum is to induce
an $\cO(\NN^2)$ {\em positive} correction to the power slope which
causes the slope (VK) condition to be satisfied and the solitons
to become stable. As demonstrated for the first time
in~\cite{NLS_NL_MS_1D,NLS_NL_MS_2D}, the size of the stability
region depends on the magnitude of the slope. This means that {\em
the transition between instability and stability is gradual rather
than sharp}, in the sense that as the soliton width $\NN$
increases from zero, the magnitude of the slope grows from zero,
hence the width of the stability region grows from zero. For
example, in the case of a Kronig-Penney lattice, the power slope
of narrow lattice solitons is exponentially small (see
Section~\ref{sec:slope}), hence the stability region is also
exponentially small. Therefore, narrow Kronig-Penney solitons are
``mathematically'' stable but ``physically'' unstable. On the
other hand, in the case of a sinusoidal lattice, the stability
region of the solitons is bigger, so that the sinusoidal lattice
solitons can be also ``physically'' stable.

In order to motivate the claims stated above, we first note that
by definition~(\ref{eq:NN}) of~$\NN$, the slope $\partial_\nu
\mathcal{P}_\nu^{(N)}$ is proportional to $\partial_\NN
\mathcal{P}_\nu^{(N)}$. Thus, the slope with respect to the
soliton width $\NN$ can be viewed as a measure for the slope with
respect to the propagation constant $\nu$. Second, we recall that
the soliton profile $u_\NN$ is an attractor for NLS solutions.
Therefore, small perturbations of the initial profile essentially
lead to small oscillations of the soliton width along the
propagation (see below). Thus, heuristically, we can view these
width oscillations as a movement along the curve
$\mathcal{P}_\nu^{(N)}$. Such movement along the curve
$P_\nu^{(N)}$ was demonstrated e.g. in Fig.~6
of~\cite{delta_pot_complete}. Since the power is conserved, a
large slope only allows for small changes of the soliton width
(i.e., stability) while a small slope allows for larger changes of
the soliton width and larger deviations from the initial state
(i.e., instability). More generally, these arguments show that
while the {\em sign} of the slope determines whether the solution
is stable or not, the {\em magnitude} of the slope $|\partial_\nu
\mathcal{P}_\nu^{(N)}|$ corresponds to the size of the stability
region. Hence, if the slope $\partial_\nu \mathcal{P}_\nu^{(N)}$
is positive {\em but small}, the stability induced by the lattice
is weak. Therefore, if the perturbation applied to the narrow
lattice soliton is large enough, the perturbation can ``overcome''
the stabilization and the solution will become unstable.

\begin{center}\begin{figure}[ht!]
\includegraphics[width=14 cm]{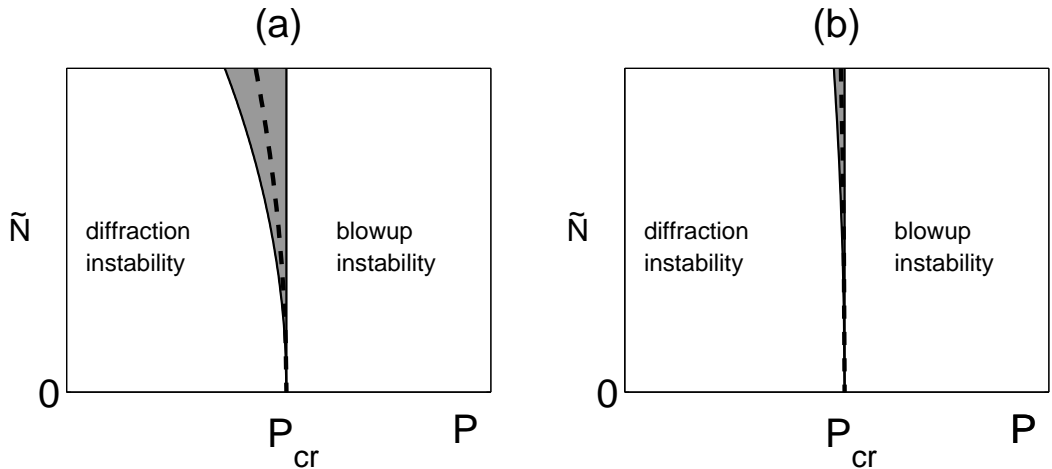}
\caption{\label{fig:stability_illustration} A schematic illustration
of stability (shaded), diffraction instability and blowup
instability regions as a function of the input beam width $\NN$ and
power $\mathcal{P}$ for narrow lattice solitons centered at a
minimum of (a) a sinusoidal lattice and (b) a Kronig-Penney lattice.
Dashed curve is $\mathcal{P}_\nu^{(N)}$.}
\end{figure}
\end{center}

A schematic illustration of the stability region in the critical
case as a function of the beam power $\mathcal{P}$ and the
relative width $\NN$ is shown in
Fig.~\ref{fig:stability_illustration}. The stability region is
centered around the lattice soliton power $\mathcal{P}_\nu^{(N)}
\cong \mathcal{P}_{cr} - \III \NN^2$, see
Eq.~(\ref{eq:un_power2_res}). By Eq.~(\ref{eq:critical_slope}) and
the above arguments, the size of the stability region depends on
the propagation constant $\nu$, the period $N$ and the lattice
$\gms({\bf x})$ only through the parameter $\NN$, and is
$\cO(\NN^2)$ small. Initial conditions to the left of the
stability region undergo a diffraction instability whereas initial
conditions to the right of the stability region undergo a blowup
instability. The separatrix between the stability region and
blowup region can be estimated by the critical power for collapse
in homogeneous medium $\mathcal{P}_{cr}$. Indeed, while the
minimal power needed for collapse depends on the beam profile, for
single-hump profiles such as $u^{(N)}_\nu$, the minimal power
needed for collapse is only slightly above
$\mathcal{P}_{cr}$~\cite{Pc}.

To illustrate these ideas numerically, we solve
Eq.~(\ref{eq:NLS_linear_lattice_dp}) for $d = d_{lat} = 2$ and $p =
3$, which correspond to the physical case of a $2D$ Kerr medium and
$\NN = 0.1$ (i.e., narrow lattice solitons). Since this is the
critical case, the lattice should have a dominant effect on the
stability (see Proposition~\ref{prop:narrow_bs_stability}). In order
to demonstrate the difference between the stabilization by the
sinusoidal lattice~(\ref{eq:V_sin_2D}) and by the Kronig-Penney
lattice~(\ref{eq:V_step_2D}), we perform a series of numerical
simulations with the initial condition $A_0(x,y) = (1 + \eps \cdot
h(x,y))u_\nu^{(N)}$. Here $\nu = \eta = 1$ and $h(x,y)$ is a random
function which is uniformly distributed in $[0,1]\times[0,1]$.
Hence, the perturbation increases the power of the initial condition
by the factor of $\approx (1 + \eps)$ with respect to the power of
the soliton $u_\nu^{(N)}$. We consider narrow solitons centered at a
lattice minimum, hence they are ``mathematically'' stable, see
Table~\ref{tab:stability}.

\begin{center}\begin{figure}[ht!]
\centerline{\includegraphics[width=0.35\textwidth]{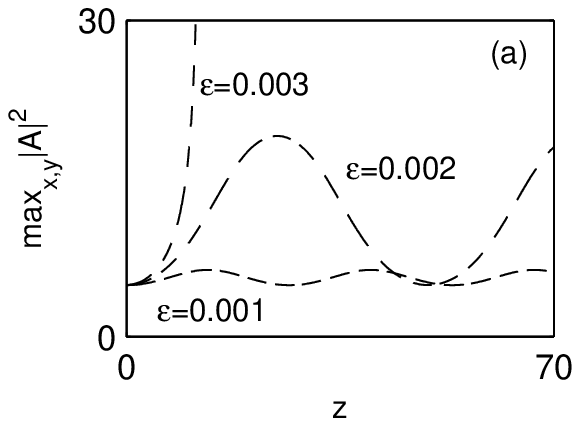}\includegraphics[width=0.3\textwidth]{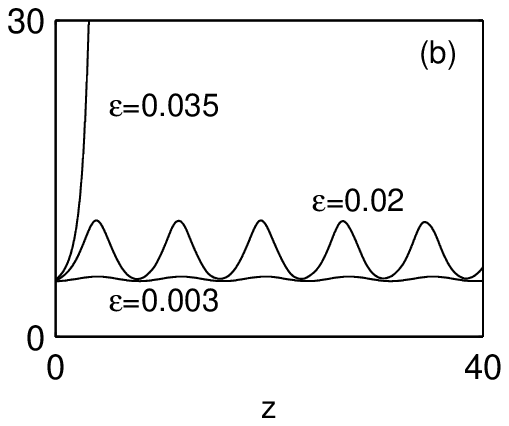}\includegraphics[width=0.3\textwidth]{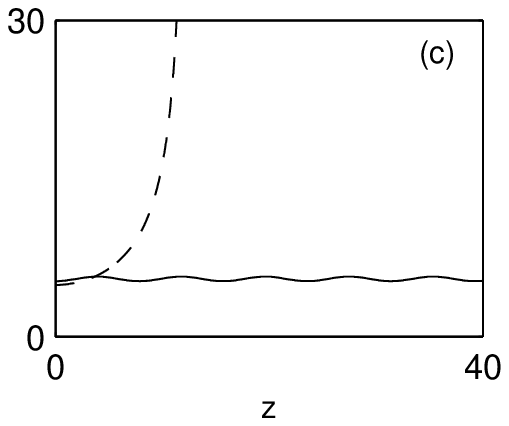}}
\caption{ Maximum intensity vs. propagation distance of narrow
lattice solitons ($N=0.1$) with power-increasing random
perturbations for (a) Kronig-Penney lattice~(\ref{eq:V_step_2D}) and
(b) sinusoidal lattice~(\ref{eq:V_sin_2D}). Comparison of the
dynamics for a sinusoidal lattice (solid) and Kronig-Penney lattices
(dashed) is shown in (c) for $\eps = 0.003$. \label{fg:increase} }
\end{figure}
\end{center}

We first note that in all the simulations in this Section, the
center of mass of the beam, which is initially perturbed from the
lattice minimum due to the random noise, remains small and close
to the lattice minimum, in accordance with Lemma~\ref{lem:com}.

In Fig.~\ref{fg:increase}(a), we show the solution for the {\em
Kronig-Penney lattice} for various values of $\eps>0$ (i.e., when
the noise increases the beam power) for $0 \le z \le 70$, i.e., over
140 diffraction lengths. For $\epsilon=0.001$ and $0.002$, the
solution undergoes a focusing-defocusing oscillations. When the
initial perturbation is further increased ($\epsilon=0.003$), the
beam undergoes collapse. The abrupt change in the dynamics between
$\epsilon=0.002$ and $\epsilon=0.003$ can be understood by looking
at the power of the beams. For the specific noise realizations in
our simulations, the power of the initial condition was slightly
{\em below} the critical power $\mathcal{P}_{cr}$ for
$\epsilon=0.001$ and $0.002$ and slightly {\em above}
$\mathcal{P}_{cr}$ for $\epsilon=0.003$. Therefore, the
beam undergoes collapse in the latter case. 

While an $\eps = 0.003$ perturbation to a Kronig-Penney lattice
soliton leads to collapse, the same perturbation applied to a narrow
{\em sinusoidal lattice} soliton only leads to small amplitude
oscillations, see Fig.~\ref{fg:increase}(b). When the perturbation
is increased to $\eps = 0.02$ the oscillations become stronger yet
the solution does not collapse. Only when the perturbation is
further increased to $\eps = 0.035$ the beam collapses in a finite
distance. As in Fig.~\ref{fg:increase}(a), we confirmed that for
$\eps = 0.003$ and $\eps = 0.02$ the beam power is below $P_{cr}$,
while for $\eps = 0.035$ it is above $P_{cr}$.

These simulations confirm that although both lattice solitons are
``mathematically'' stable, sufficiently large perturbations can
still cause these stable solitons to undergo collapse~\footnote{Note
that the typical perturbations in experimental setups are at least
of few percents.}. This demonstrates that collapse and stability can
co-exist, see also~\cite{FM:01,FW:03}. Moreover, these simulations
also support the heuristic argument presented in
Section~\ref{sub:slope} that the upper boundary of the stability
region can be estimated by the critical power for collapse in
homogeneous medium~$P_{cr}$.

\begin{center}\begin{figure}
\centerline{
\includegraphics[width=0.6\textwidth]{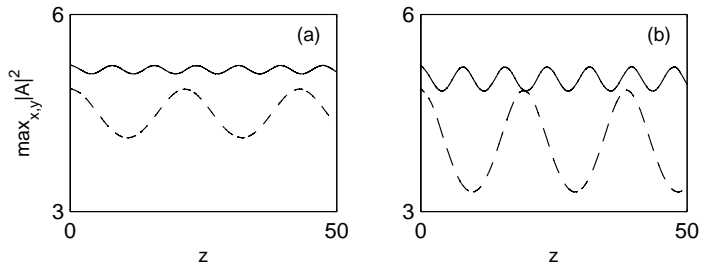}
} \caption{ Maximum intensity vs. propagation distance of narrow
($N=0.1$) lattice solitons with power-decreasing random
perturbations for sinusoidal (solid) and Kronig-Penney lattices
(dashed) with (a) $\eps = - 0.001$ and (b) $\eps = - 0.003$.
\label{fg:decrease}}
\end{figure}
\end{center}

In Fig.~\ref{fg:decrease}, we show the solutions for $\eps = -
0.001$ and $\eps = - 0.003$ (i.e., when the noise decreases the beam
power). The comparison between the two lattices for the same value
of $\epsilon$ shows that the stabilization by the sinusoidal lattice
is much stronger than by a Kronig-Penney lattice. Additional
simulations (data not shown) show that the difference between the
stabilization by the two lattices becomes more pronounced as $N$
becomes smaller. Indeed, for a Kronig-Penney lattice, the boundaries
of the lattice are located far in the soliton tail region. Thus,
their presence can prevent broadening only once the narrow beam has
undergone significant broadening. On the other hand, a sinusoidal
lattice acts at any position in the central region of the soliton,
hence, it has a much more pronounced effect.

The results shown in Figs.~\ref{fg:increase} and~\ref{fg:decrease}
confirm that Kronig-Penney lattice solitons are ``physically
unstable'' (i.e., an extremely small stability region) whereas
sinusoidal lattice solitons can be ``physically stable''
(not-so-small stability region). Indeed, a comparison between
these two lattices for the same value of $\epsilon$ shows that for
narrow lattice solitons, {\em the same perturbation leads to
collapse in the case of a Kronig-Penney lattice but only to small
oscillations and stable behaviour in the case of a sinusoidal
lattice}, see Fig.~\ref{fg:increase}(c) and
Fig.~\ref{fg:decrease}.

\subsection{``Mathematical'' vs. ``physical''
instability}\label{sub:drift} We now consider narrow lattice
solitons centered at a lattice maximum. According to
Proposition~\ref{prop:narrow_bs_stability}, these solitons are
unstable as they violate the spectral condition. Indeed, we showed
that these solitons undergo a drift instability away from the
lattice maximum. Since there is no drift for $\lambda_{0,j} = 0$, by
continuity, the drift rate should be ``small'' for small negative
values of $\lambda_0^{(N)}$. Indeed, combining
Eqs.~(\ref{eq:ev_expansion_res}) and~(\ref{eq:com_dyn_sol}), one
sees that for $v_{jj} < 0$,
\begin{equation}\label{eq:com_dyn_sol_lambda} \fl
\Av{x_j(z)} \sim \Av{x_j(0)} \cosh(\Omega z) +
\frac{\Av{\dot{x}_j(0)}}{\Omega} \sinh(\Omega z), \quad \Omega =
2\sqrt{\frac{\eta d |\lambda_0^{(N)}|}{\delta}}.
\end{equation}
Thus, if $-\lambda_0^{(N)}$ is small, the instability develops very
slowly. In this case, the solitons are ``mathematically'' unstable
but can be ``physically stable'', i.e., the instability does not
develop over the propagation distance of the experiment. If, on the
other hand, the instability does develop over such distances, one
can say that the soliton is also ``physically unstable''.

\begin{figure} \centerline{
\includegraphics[width=0.315\textwidth]{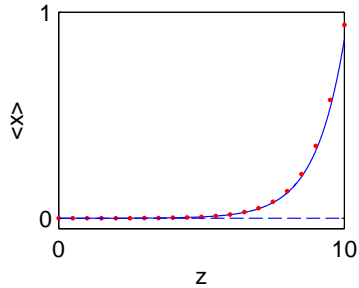}
} \caption{ Center of mass of the solution of
Eq.~(\ref{eq:NLS_linear_lattice_dp}) with $d=1$, $p=3$ and a
sinusoidal lattice~(\ref{eq:1D_sin}) (solid line) and a KP
lattice~(\ref{eq:1D_KP}) (dashed line). The lattice parameters are
$N=0.07$ and $V_0 = 2.5$; the initial shift of the soliton center
is $\delta_c = 10^{-4}$. The analytical
formula~(\ref{eq:com_dyn_sol_lambda_example}) (red dots) is nearly
indistinguishable from the numerical result. \label{fg:drift}}
\end{figure}

In order to demonstrate the drift instability associated with
violation of the spectral condition, and in particular, the
importance of the magnitude of $\lambda_0^{(N)}$, we solve
Eq.~(\ref{eq:NLS_linear_lattice_dp}) with $d=1$ and $p=3$ for a
sinusoidal lattice
\begin{equation}\label{eq:1D_sin}
\gms(Nx) = V_0 \cos(2 \pi N x),
\end{equation}
and also for a Kronig-Penney lattice with the unit cell that
consists of a periodic array of cells of size $1/N$, where for each
cell,
\begin{equation}
\gms (Nx) = \left\{
\begin{array}{ll}
V_0, & |x| < \frac{1}{4N}\\
0, & \frac{1}{4N} < |x| < \frac{1}{2N}.
\end{array}
\right. \label{eq:1D_KP}
\end{equation}
We excite the instability by shifting the soliton center slightly
off the lattice maximum, i.e., we use the initial condition $A_0(x)
= u_\nu^{(N)}(x - \delta_c)$. In Fig.~\ref{fg:drift} we show the
center of mass of the solution for $N=0.07$, $\nu = 10$, $V_0 = 2.5$
and $\delta_c = 10^{-4}$. For these parameters, $\Av{x(0)} =
\delta_c$ and $\Av{\dot{x}(0)} = 0$ so that by
Eq.~(\ref{eq:com_dyn_sol_lambda}),
\begin{equation}\label{eq:com_dyn_sol_lambda_example}
\Av{x_j(z)} \sim \delta_c \cosh(\Omega z), \quad \quad \Omega =
2\sqrt{\frac{\eta d \big|\lambda_0^{(N)}\big|}{\delta}}.
\end{equation}
This exponential drift-rate is indeed observed in the simulation for
the sinusoidal lattice soliton, see Fig.~\ref{fg:drift}. This shows
that while the sign of $\lambda_0^{(N)}$ determines whether the
soliton is (``mathematically'') stable or unstable, the magnitude of
$|\lambda_0^{(N)}|$ determines the rate of the instability dynamics.

The drift rate for the KP lattice soliton is several orders of
magnitude smaller than for the sinusoidal lattice soliton.
Intuitively, this is because unlike the sinusoidal lattice, the KP
lattice affects the soliton profile (and hence the dynamics) only in
the soliton tail region. As expected, the magnitude of
$\lambda_0^{(N)}$ is much larger for the sinusoidal lattice soliton
($\lambda_0^{(N)} \cong -0.05$) than for the KP lattice soliton
($\lambda_0^{(N)} \cong -2\cdot10^{-5}$). Moreover, the drift rate
of the KP lattice soliton is considerably smaller than the one
predicted by Eq.~(\ref{eq:com_dyn_sol_lambda_example}) with
$\lambda_0^{(N)} \cong -2\cdot10^{-5}$. This ``mismatch'' is not
surprising, since Eq.~(\ref{eq:com_dyn_sol_lambda_example}) is not
valid for the KP lattice, see also Section~\ref{subsec:lat_type}.

\begin{figure}
\centerline{
\includegraphics[width=0.9\textwidth]{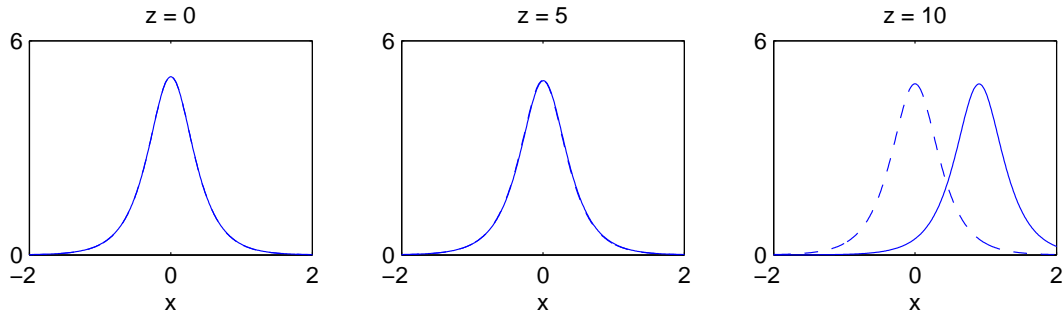}
} \caption{Beam profiles at several propagation distances for the
data of Fig.~\ref{fg:drift}. The beam profiles for the sinusoidal
lattice~(\ref{eq:1D_sin}) (solid line) and the KP
lattice~(\ref{eq:1D_KP}) (dashed line) at $z=0$ and $z = 5$ are
indistinguishable. \label{fg:drift_cs}}
\end{figure}

At a propagation distance of $z = 5$, both the sinusoidal and the KP
lattice solitons hardly shift from their initial location, see
Fig.~\ref{fg:drift_cs}. At a propagation distance of $z = 10$,
however, the sinusoidal lattice soliton drifts more than one soliton
width whereas the Kronig-Penney lattice soliton hardly drifts at
all. In that sense, since the propagation distance in the
simulations corresponds to a distance of $20$ diffraction lengths,
which is longer than most devices in optics, the ``mathematically
unstable'' KP soliton is ``physically stable''.

\section{Discussion and comparison with previous studies}\label{sec:summary}
Most rigorous studies on stability and instability of lattice
solitons are based on the Grillakis, Shatah and Strauss (GSS)
theory~\cite{GSS,GSS2}. Let $u_\nu^{(N)}>0$, let
\begin{equation*}\label{eq:H+P}
d(\nu) = \mathcal{H} + \nu \mathcal{P} = \int \left[ |\nabla
u_\nu^{(N)}|^2 + \left(V(N{\bf x}_{lat}) + \nu
\right)\left(u_\nu^{(N)}\right)^2 - \frac{2}{p+1}
\left(u_\nu^{(N)}\right)^{p+1}\right]d{\bf x},
\end{equation*}
let $p(d'') = 1$ if $d'' > 0$ and $p(d'') = 0$ if $d'' < 0$, and let
$n_-(L_{+,\nu}^{(N)})$ be the number of negative eigenvalues of the
operator $L_{+,\nu}^{(N)}$. Then, $u_\nu^{(N)}e^{i \nu z}$ is
orbitally stable if $n_-(L_{+,\nu}^{(N)}) = p(d'')$, and orbitally
unstable if $n_-(L_{+,\nu}^{(N)}) - p(d'')$ is odd~\cite{GSS,GSS2}.
For example, stability of lattice solitons was studied
in~\cite{Berge-parabolic,Reika-01-harmonic,Reika-03-gen-pot,Reika-05-gen-pot}
using the GSS theory. In addition, after this paper was submitted,
we found out that the GSS theory was applied to narrow lattice
solitons in the critical case by Lin and Wei~\cite{Lin_Wei}.

Since $d'(\nu) = \int \left(u_\nu^{(N)}\right)^2 d{\bf x}$, the sign
of $d''$ is the same as the sign of the power slope. Hence, in the
GSS theory stability and instability depend on a combination of the
slope condition and a spectral condition: If both the slope
condition and the spectral condition are satisfied, the soliton is
stable, whereas if either the slope condition is satisfied and
$n_-(L_{+,\nu}^{(N)})$ is even, or if the slope condition is
violated and $n_-(L_{+,\nu}^{(N)})$ is odd, the soliton is unstable.
There are two cases not covered by the GSS theory: When the slope
condition is satisfied and $n_-(L_{+,\nu}^{(N)})$ is odd, and when
the slope condition is violated and $n_-(L_{+,\nu}^{(N)})$ is even.
As Theorem~\ref{theo:stability-nls} shows, in both cases the
solitons are unstable. Hence, there is a ``decoupling'' of the slope
and spectral conditions, in the sense that both are needed for
stability, and violation of either of them would lead to
instability.

In~\cite{NLS_NL_MS_1D,NLS_NL_MS_2D,delta_pot_complete} it was {\em
observed numerically} that violation of the slope condition leads to
a width instability, whereas violation of the spectral condition
leads to a drift instability. Unlike these
studies, in this study we 
{\em prove} that violation of the spectral condition leads to a
drift instability. Moreover, we show that a drift instability occurs
in any direction $x_j$ for which the corresponding eigenvalue
$\lambda_{0,j}^{(N)}$ is negative, and that the drift rate is
determined by the magnitude of $\lambda_{0,j}^{(N)}$.~\footnote{A
generalization of these results to non-narrow beams can be found
in~\cite{drift_paper}. } This further shows that violation of the
spectral condition leads to an instability, regardless of the slope
condition and of whether $n_-(L_{+,\nu}^{(N)})$ is even or odd.



In previous studies it was also observed that in the subcritical
case, lattice solitons centered at a lattice minimum of all widths
are stable. In the critical case, it was shown that lattice solitons
are stable only if they are narrower than a few lattice periods, see
e.g.,~\cite{efrem-prl2003,ZiadY}. These results are in agreement
with Table~\ref{tab:stability} in the subcritical and critical
cases, and imply that our analytical results are valid beyond the
regime of narrow lattice solitons.
In~\cite{mihal-pra2005,mihal-3d-optics} it was also shown that in
the supercritical case, the lattice can stabilize sufficiently wide
lattice solitons centered at a lattice minimum but cannot stabilize
narrow lattice solitons, in agreement with our results. Note,
however, that unlike most previous works, our results are valid for
any dimension~$d$, lattice dimension~$d_{lat}$ and nonlinearity
exponent~$p$.

Another difference from previous studies on linear lattices is that
we introduce a quantitative approach to the notions of stability and
instability. Thus, we show that the strength of radial stabilization
depends on the magnitude of the slope. Hence, in the critical case,
the stability of the soliton is ``mathematical'' but not
``physical''. Similarly, we show that the strength of the transverse
instability depends on the value of the perturbed zero eigenvalue
$\lambda_0^{(N)}$. Hence, for narrow solitons centered at a lattice
maximum, the instability is ``mathematical'' but not necessarily
``physical''. In such cases, the stabilization/destabilization of
narrow lattice solitons is highly sensitive to the lattice details.
This sensitivity becomes smaller as the soliton width increases, and
is of considerably less importance for $\cO(1)$ solitons, which is
probably why this feature was not observed in previous studies.



\ack We thank S. Toledo for his help with numerical computations of
the eigenvalues. The research of Gadi Fibich and Yonatan Sivan was
partially supported by grant 2006-262 from the United States–-Israel
Binational Science Foundation (BSF), Jerusalem, Israel. N.K.E.
acknowledges funding by the European Social Fund (75\%) and National
Resources (25\%)-Operational Program for Educational and Vocational
Training II (EPEAEK II) and particularly the Program PYTHAGORAS.

\appendix

\section{Proof of Lemma~\ref{lem:sol}} \label{ap:nls-proof}
The approach used here is similar to~\cite{NLS_NL_MS_1D,FW:03}.
Substituting the expansion~(\ref{eq:lattice-series}) in
Eq.~(\ref{eq:uNN-eqn}) gives
\begin{equation}
\nabla^2 u_\NN + u_\NN^p - \Big(1 + \NN^2 \tilde \gms_2(\tilde
{\bf x}_{lat}) \Big) u_\NN + \cO(\NN^4) = 0. \label{}\nn
\end{equation}
Let $u_\NN(\tilde {\bf x})$ be given by Eq.~(\ref{eq:expansion}).
Then, the equation for $g$ is
\begin{eqnarray}
\nabla^2 g(\tilde {\bf x}) + p\UU^{p-1}g - \nu g = \tilde
\gms_2(\tilde {\bf x}_{lat})\UU(|\tilde {\bf x}|). \nn
\end{eqnarray}
Therefore,
\begin{eqnarray}
g(\tilde {\bf x}) = - L_+^{-1} \big[\tilde \gms_2(\tilde {\bf
x}_{lat}) \UU(|\tilde {\bf x}|) \big]. \label{eq:g-eqn}
\end{eqnarray}

\section{Proof of Lemma~\ref{lem:power}} \label{app:power-proof}
By Eq.~(\ref{eq:uNN-sol}), the power of the rescaled lattice soliton
$\mathcal{P}_\NN = \int \big(u_\NN(\tilde {\bf x}) \big)^2 d\tilde
{\bf x}$ is given by
\begin{eqnarray}
\mathcal{P}_\NN &=& \mathcal{P}_{\nu=1} - 2 \NN^2 \int \UU(\tilde
r)
L_+^{-1} \big[\tilde \gms_2(\tilde {\bf x}_{lat}) \UU \big] d\tilde {\bf x} + \cO(\NN^4) \nn \\
&=& \mathcal{P}_{\nu=1} - 2 \NN^2 \int \tilde \gms_2(\tilde {\bf
x}_{lat}) \UU(\tilde r) L_+^{-1} \left[\UU\right] d\tilde {\bf x}
+ \cO(\NN^4), \label{eq:un_power1}
\end{eqnarray}
where $\mathcal{P}_{\nu=1} = \int \UU^2(\tilde r) d\tilde {\bf x}$
and $\tilde r = |\tilde {\bf x}|$. In order to proceed, we prove
the following Lemma:
\begin{lem}\label{lem:L+}
Let $\UU_\eta$ be the solution of Eq.~(\ref{eq:UU-eta-eqn}) and
let $L_{+,\eta}$ be given by Eq.~(\ref{eq:L+nu}). Then,
$L_{+,\eta}^{-1} \UU_\eta = - \D_\eta \UU_\eta$.
\end{lem}

\nit{\bf Proof:} Differentiating Eq.~(\ref{eq:UU-eta-eqn}) with
respect to $\eta$ gives
\begin{eqnarray}
&& \D_\eta\ (\nabla^2 \UU_\eta) + \D_\eta\ \UU_\eta^{p} - \D_\eta\
(\eta \UU_\eta) = \nabla^2\ (\D_\eta \UU_\eta) + p\UU_\eta^{p-1}\
(\D_\eta \UU_\eta) -\UU_\eta - \eta \D_\eta \UU_\eta = \nn \\ && =
-L_+ \D_\eta \UU_\eta - \UU_\eta = 0. \quad \Box. \nn
\end{eqnarray}
Since $\UU_\eta(\tilde r) =
\eta^{\frac{1}{p-1}}\UU(\sqrt{\eta}\tilde r)$, then
\begin{eqnarray}
\D_\eta \UU_\eta = \frac{1}{p-1} \eta^{\frac{1}{p-1}-1} \UU +
\eta^{\frac{1}{p-1}} \left(\frac{1}{2} \eta^{-\frac{1}{2}} \tilde
r \right) \UU_{\tilde r}. \nn
\end{eqnarray}
Therefore, $L_+^{-1} \UU = L_+^{-1} \UU_{\eta=1} = - (\D_\eta
\UU_\eta)_{\eta = 1} = - \frac{1}{p-1} \UU - \frac{1}{2} \tilde r
\UU_{\tilde r}$. Substituting in Eq.~(\ref{eq:un_power1}) gives
\begin{eqnarray}\label{eq:PNN}
\mathcal{P}_\NN &=& \mathcal{P}_{\nu=1} + 2 \NN^2 \int \tilde \gms_2(\tilde {\bf x}_{lat}) \UU \left( \frac{\UU}{p-1} + \tilde r \frac{\UU_{\tilde r}}{2} \right)d \tilde {\bf x} + \cO(\NN^4).
\end{eqnarray}
Since $\tilde V_2$ is given by Eq.~(\ref{eq:v2_jj}),
Eq.~(\ref{eq:PNN}) can be written as
\begin{eqnarray}
\mathcal{P}_\NN &=& \mathcal{P}_{\nu=1} - \III \NN^2
\sum_{j=1}^{d_{lat}} v_{jj} + \cO(\NN^4), \label{eq:un_power2} \nn
\end{eqnarray}
where $\III$ is given by
\begin{eqnarray}\label{eq:C_V_app}
\III &=& - \int \tilde x_j^2 \left( \frac{2\UU^2}{p-1} + \tilde r
\UU \UU_{\tilde r} \right)d \tilde {\bf x}.
\end{eqnarray}
To bring $\III$ to the form~(\ref{eq:C_V}), we note that
$$
\nabla \cdot \left( b(\tilde r) \tilde {\bf x} \right) =
\frac{1}{\tilde r^{d-1}} \frac{\partial}{\partial \tilde r}
\left(\tilde r^d b(\tilde r) \right) = \frac{1}{\tilde r^{d-1}}
\left( d\tilde r^{d-1} b + \tilde r^d b'\right) = d b + \tilde r
b'.
$$
Substituting $b(\tilde r) = \tilde r^{\frac{4}{p-1} - d}
\UU^2(\tilde r)$ shows that
\begin{eqnarray}
\nabla \cdot \left( \tilde r^{\frac{4}{p-1} - d} \UU^2(\tilde r)
{\bf x} \right) &=& d \tilde r^{\frac{4}{p-1} - d} \UU^2 +
\left(\frac{4}{p-1} - d\right)
\tilde r^{\frac{4}{p-1} - d} \UU^2 + 2 \tilde r^{\frac{4}{p-1} - d + 1} \UU \UU' \nn \\
&=& 2 \tilde r^{\frac{4}{p-1} - d} \left( \frac{2\UU^2}{p-1} +
\tilde r \UU \UU_{\tilde r} \right). \nn
\end{eqnarray}
Thus, we can rewrite Eq.~(\ref{eq:C_V_app}) as
\begin{eqnarray}
\III &=& - \frac{1}{2} \int \frac{\tilde x_j^2}{\tilde
r^{\frac{4}{p-1} - d}} \nabla \cdot \left( \tilde r^{\frac{4}{p-1}
- d} \UU^2 \tilde {\bf x} \right)
d\tilde {\bf x} \label{eq:III3} \\
&=& \frac{1}{2} \int r^{\frac{4}{p-1} - d} \UU^2 \tilde {\bf x}
\cdot \nabla \left( \frac{\tilde x_j^2}{\tilde r^{\frac{4}{p-1} -
d}} \right) d\tilde {\bf x} \nn
\\
&=& \frac{1}{2} \int r^{\frac{4}{p-1} - d} \UU^2 \tilde {\bf x}
\cdot \left( \frac{2 \tilde x_j \hat{\bf e}_{\tilde x_j}}{\tilde
r^{\frac{4}{p-1} - d}} - \left(\frac{4}{p-1} - d\right)
\frac{\tilde x_j^2\hat{\bf e}_{\tilde r}}{\tilde r^{\frac{4}{p-1}
- d + 1}} \right) d\tilde
{\bf x} \nn \\
&=& \frac{1}{2} \int \UU^2 \left(2 \tilde x^2_j -
\left(\frac{4}{p-1} - d\right) \tilde x_j^2 \right) d \tilde {\bf
x} = \frac{1}{2d} \int \tilde r^2 \UU^2 \left( 2 - \frac{4}{p-1} +
d \right) d\tilde {\bf x}. \nn
\end{eqnarray}
Finally, by the dilation transformation~(\ref{eq:eta}),
\begin{eqnarray}
\mathcal{P}_\nu^{(N)} &\equiv& \int \left(u_\nu^{(N)}({\bf x})
\right)^2 d{\bf x} = \eta^{\frac{2}{p-1}} \int \left(u_\NN(\tilde
{\bf x}) \right)^2 d{\bf x} = \eta^{\frac{4-d(p-1)}{2(p-1)}}
\mathcal{P}_\NN. \nn
\end{eqnarray}

\section{Proof of Lemma~\ref{lem:ev}}\label{app:ev}
Consider the eigenvalue problem
\begin{eqnarray}\label{eq:ev_problem}
L_{+,\nu}^{(N)} f^{(N)}_{\nu,j}({\bf x}) = \lambda_{0,j}^{(N)}
f^{(N)}_{\nu,j}.
\end{eqnarray}
Multiplying Eq.~(\ref{eq:ev_problem}) by $ f^{(N)}_{\nu,j}$ and
integrating gives
\begin{eqnarray}\label{eq:f_nu_N_L2}
\int f^{(N)}_{\nu,j} L_{+,\nu}^{(N)} f^{(N)}_{\nu,j} d{\bf x} =
\lambda_{0,j}^{(N)} \int \left( f^{(N)}_{\nu,j}\right)^2 d{\bf x}.
\end{eqnarray}
We recall that in the absence of a lattice, the operator
$L_{+,\nu}^{(N)}$ reduces to $L_{+,\nu}$, see Eq.~(\ref{eq:L+nu}),
which has $d$~zero eigenvalues $\lambda_{0,j} = 0$, with the
corresponding eigenfunctions $\frac{\partial \UU_\nu}{\partial x_j}$
for $j = 1,\dots,d$, see Eq.~(\ref{eq:wj0}). By
Eq.~(\ref{eq:u-nu-N-sol}), in the presence of the lattice,
$u_\nu^{(N)} = \UU_\eta + \eta^{\frac{1}{p-1}}\cO(\NN^2)$.
Similarly, by Eq.~(\ref{eq:scaling_add}),
Eq.~(\ref{eq:lattice-series}) and Eq.~(\ref{eq:v2_jj}), we can
expand the potential as
\begin{eqnarray}\label{eq:VN_NN_ap} \fl
V(N {\bf x}_{lat}) = V(\NN \tilde {\bf x}_{lat}) = V(0) + \eta
\tilde V(\NN \tilde {\bf x}_{lat}) = V(0) + \eta \left(\NN^2
\tilde V_2(\tilde {\bf x}_{lat}) + \cO(\NN^4) \right) \\
= V(0) + N^2 \sum_{j=1}^{d_{lat}} v_{jj} \tilde x_j^2 + \eta \cdot
\cO(\NN^4) = V(0) + \eta N^2 \sum_{j=1}^{d_{lat}} v_{jj} x^2_j +
\eta \cdot \cO(\NN^4). \nonumber
\end{eqnarray}
Consequently, the operator $L^{(N)}_{+,\nu}$ can be expanded as
\begin{eqnarray}\label{eq:L_+_nu_exp}
L_{+,\nu}^{(N)} &=& -\nabla^2 - p \left(u_\nu^{(N)}\right)^{p-1} +
\nu + V(N {\bf x}_{lat}) \\
&=& -\nabla^2 - p\left(\UU_\eta + \eta^{\frac{1}{p-1}}\cO(\NN^2)
\right)^{p-1} + \nu +
V(0) + \cO(N^2) \nn \\
&=& -\nabla^2 - p\UU_\eta^{p-1} + \eta + \cO(N^2) = L_{+,\eta} +
\cO(N^2). \nn
\end{eqnarray}
Therefore, we expand
\begin{eqnarray}\label{eq:ef1}
f^{(N)}_{\nu,j}({\bf x}) = \frac{\partial \UU_\eta}{\partial x_j}
\left(1 + \cO(N^2)\right), \quad \quad \lambda_{0,j}^{(N)} =
\delta_j N^2 + \cO(N^4).
\end{eqnarray}
By Eqs.~(\ref{eq:u-nu-N-sol}) and~(\ref{eq:ef1}), we can also
rewrite the eigenfunction $f^{(N)}_{\nu,j}$ as
\begin{eqnarray}\label{eq:ef2}
f^{(N)}_{\nu,j}({\bf x}) = \frac{\partial u_\nu^{(N)}}{\partial
x_j}\left(1 + \cO(N^2)\right).
\end{eqnarray}
We now use the approximations~(\ref{eq:ef1}) and~(\ref{eq:ef2}) in
order to evaluate the terms in Eq~(\ref{eq:f_nu_N_L2}). By
Eq.~(\ref{eq:ef1}), the right-hand-side of
Eq.~(\ref{eq:f_nu_N_L2}) is equal to
\begin{eqnarray}\label{eq:ev_problem_1}
\lambda_{0,j}^{(N)} \int \left(f_\nu^{(N)}\right)^2 d{\bf x} &=&
\left(N^2 \delta_j + \cO(N^4)\right) \left(\int \left(
\frac{\partial \UU_\eta}{\partial x_j} \right)^2  d{\bf x} + \cO(N^2)\right) \nn \\
&=& N^2 \delta_j \int \left( \frac{\partial \UU_\eta}{\partial
x_j} \right)^2  d{\bf x} + \cO(N^4).
\end{eqnarray}
By Eq.~(\ref{eq:ef2}) the left-hand-side of
Eq.~(\ref{eq:f_nu_N_L2}), approximation~(\ref{eq:ef2}) is equal to
\begin{eqnarray}\label{eq:ef_norm_2}
\int f^{(N)}_{\nu,j} L_{+,\nu}^{(N)} f^{(N)}_{\nu,j} d{\bf x} =
\int \frac{\partial u_\nu^{(N)}}{\partial x_j} L_{+,\nu}^{(N)}
\frac{\partial u_\nu^{(N)}}{\partial x_j}  d{\bf x} + \cO(N^4),
\end{eqnarray}
where the error term is $\cO(N^4)$ due to the properties of the
Rayleigh quotient, see e.g.,~\cite{Trefethen}.

The integral term on the right-hand-side of
Eq.~(\ref{eq:ef_norm_2}) is equal to
\begin{equation}\label{eq:1}
\int \frac{\partial u_\nu^{(N)}}{\partial x_j} L_{+,\nu}^{(N)}
\frac{\partial u_\nu^{(N)}}{\partial x_j} d{\bf x} = \frac{1}{2}
\int \left( u_\nu^{(N)} \right)^2 \frac{\partial^2}{\partial
x^2_j} V(N {\bf x}_{lat}) d{\bf x} .
\end{equation}
Indeed, differentiating Eq.~(\ref{eq:uN-eqn}) with respect to
$x_j$ gives
\begin{equation}\label{eq:0}
L_{+,\nu}^{(N)} \frac{\partial u_\nu^{(N)}}{\partial x_j} = -
\left( \frac{\partial V(N {\bf x}_{lat})}{\partial x_j} \right)
u_\nu^{(N)}.
\end{equation}
Multiplying Eq.~(\ref{eq:0}) by $\frac{\partial}{\partial x_j}
u_\nu^{(N)}$, integrating over ${\bf x}$ and integrating by parts
gives Eq.~(\ref{eq:1}). Using Eq.~(\ref{eq:VN_NN_ap}), the
right-hand-side of Eq.~(\ref{eq:1}) is given by
\begin{equation}\label{eq:3}
\frac{1}{2} \int \left( u_\nu^{(N)} \right)^2
\frac{\partial^2}{\partial x^2_j} V(N {\bf x}_{lat}) d{\bf x} =
\eta N^2 v_{jj} \int \UU_\eta^2  d{\bf x} + \cO(N^4).
\end{equation}
Comparing the approximation~(\ref{eq:ev_problem_1}) for the
left-hand-side of Eq.~(\ref{eq:f_nu_N_L2}) with the
approximation~(\ref{eq:3}) for the right-hand-side of
Eq.~(\ref{eq:f_nu_N_L2}) shows that 
\begin{equation}\label{eq:4}
\delta_j \int \left( \frac{\partial \UU_\eta}{\partial x_j}
\right)^2
= \eta v_{jj} \int \UU_\eta^2.
\end{equation}
Hence,
\begin{equation}
\delta_j = \eta v_{jj}\frac{\int \UU_\eta^2}{\int \left(
\frac{\partial \UU_\eta}{\partial x_j} \right)^2} = d v_{jj}
\frac{\int \UU^2}{\int {\UU'}^2}.
\end{equation}
Similar results were obtained in~\cite{Lin_Wei} for a soliton
centered at a general non-degenerate critical point of the lattice
(i.e., without assuming that the critical point is symmetric with
respect to ${\bf x}_{lat}^{(0)}$).

By the Pohozaev identities for Eq.~(\ref{eq:UU-eqn})
(see~\cite{Sulem-99}, pp.~76), $\frac{\int \UU^2}{\int {\UU'}^2} =
\frac{p(2-d) + 2 + d}{d(p-1)} \equiv \frac{\delta}{d}$. Therefore,
we get that
\begin{equation}
\delta_j = \delta v_{jj}.
\end{equation}

\section{Computing small eigenvalues of a very large matrix}\label{app:code}
When $d \ge 2$, the discretized operator $L_+^{(N)}$ is
represented by an extremely large matrix. Hence, straightforward
application of standard numerical routines (such as Matlab's {\tt
eig/eigs}) usually either fails to give accurate results or does
not converge.

In order to overcome this numerical problem, we used a more
efficient and robust numerical method based on the Arnoldi
algorithm (performed by ARPACK~\cite{arpack}, which is available
in Matlab through the function {\tt eigs}). Essentially, we
compute the largest-magnitude eigenvalues of the inverse matrix
$A^{-1}$ which correspond to the smallest eigenvalues of the
matrix~$A$.

We compute the $LU$ factorization of $A$ with complete pivoting.
Then, we shift the values on the main diagonal of $U$ by a small
value in order to avoid numerical errors that might result from
singularity of the matrix during the computation of $A^{-1}$.
Then, in order to avoid working with the explicit from of the
inverse matrix $A^{-1}$ which is dense, we compute $A^{-1}$
implicitly through the subfunction ${\tt LUPinv}$ and apply it to
the function ${\tt eigs}$. This way, we exploit the sparsity of
the $LU$ factorized matrices $U$ and $L$. The function ${\tt
eigs}$ then computes the desired number of eigenvalues of largest
magnitude.

The following code was given to us by Prof. S. Toledo:
\bigskip

\noindent {\tt function [V,d] = ev\_calculation(A,ev\_number,eps)

[m n] = size(A); normA = norm(A,1);

[L,U,P,Q] = lu(A,1.0);

for j=1:n

    \quad if (abs(U(j,j)) < eps*normA)

    \quad \quad    U(j,j) = eps*normA;

    \quad end

end

h = @LUPinv;

opts.issym = true;

opts.isreal = true;

opts.tol = eps;

[V,D] = eigs(h,n,ev\_number,'LM',opts);

\bigskip
function Y = LUPinv(X)

\quad    Y1 = P*X;

\quad    Y2 = L $\setminus$ Y1;

\quad    Y3 = U $\setminus$ Y2;

\quad    Y  = Q*Y3;

end

\noindent end }

\section{Proof of Lemma~\ref{lem:com}} \label{app:accel}
Multiplying Eq.~(\ref{eq:NLS_linear_lattice_dp}) by $A^*$ and
subtracting the conjugate equation gives
\begin{equation}
\frac{d}{dz}|A|^2 = i A^* \nabla^2 A + c.c.,\nn
\end{equation}
where c.c. stands for complex conjugate. Multiplying by ${\bf x}$
and integrating over ${\bf x}$ gives
\begin{eqnarray}
\frac{d}{dz} \int {\bf x} |A|^2 
&=& \int i {\bf x} A^* \nabla^2 A + c.c. = - i \int \nabla A (dA^* +
{\bf x} \cdot \nabla A^*) + c.c. \nonumber \\
&=& 2d\ Im \int A^* \nabla A. \label{eq:velocity_p}
\end{eqnarray}
Differentiating Eq.~(\ref{eq:velocity_p}) yields
\begin{eqnarray}
\frac{d^2}{dz^2} \int {\bf x} |A|^2 &=& 2d\ Im \int (A_{z}^*
\nabla A + A^* \nabla A_z)
\nn \\
&=& 2d\ Im \int (A_{z}^* \nabla A - A_z \nabla A^*) = 4d\ Im \int
A_{z}^* \nabla A \nn
\\ &=& - 4d\ Re \int \left(\nabla^2 A^* + |A|^{p-1} A^* -
\gms(Nx) A^* \right)\nabla A. \nn
\end{eqnarray}
The first two terms vanish since they are complete derivatives.
Therefore,
\begin{eqnarray}
\frac{d^2}{dz^2} \int {\bf x} |A|^2 &=& 4d\ Re \int \gms(Nx)
A^* \nabla A \nn \\
&=& 2d \int \gms(N{\bf x}_{lat} ) \nabla |A|^2 = - 2d \int |A|^2
\nabla \gms(N{\bf x}_{lat} ).
\end{eqnarray}
Finally, by Eq.~(\ref{eq:VN_NN}),
\begin{eqnarray}\label{eq:accel}
\frac{d^2}{dz^2} \int x_j |A|^2 = - 4 N^2 d \eta v_{jj} \int x_j
|A|^2 + \cO(N^4).
\end{eqnarray}

\section*{References}
\bibliographystyle{unsrt}

\end{document}